\documentstyle[epsf]{article}
\newcommand{\subsetX}{{\subset \atop X}}
\newcommand{\bookfig}[5]{
\begin{figure}\centering\fbox{\epsfysize=#5cm \epsfbox{#1}} 
\caption[#2]{\small #4}\label{#3}
\end{figure}}
\begin{document}
\title{On Overlapping Divergences}
\author{Dirk KREIMER
\thanks{Heisenberg Fellow, email: dirk.kreimer@uni-mainz.de}\\
Dept.~of Physics\\
Mainz Univ.\\
55099 Mainz\\
Germany
}
\date{\small \bf MZ-TH/98-35, hep-th/9810022}
\maketitle
\begin{abstract}
Using set-theoretic considerations, 
we show that the forest formula for overlapping divergences
comes from the Hopf algebra of rooted trees.
\end{abstract}
\section*{Motivation and Introduction}
The process of renormalization is governed by the forest formula, as
derived for example in \cite{Zimm}.
The underlying combinatorics is directly related to the Hopf algebra
structure of rooted trees. This is evident in the case of Feynman
diagrams which only provide nested or disjoint subdivergences.
It is the purpose of this paper to show that the same Hopf algebra
appears in the study of overlapping divergences. This was already shown
using Schwinger Dyson equations \cite{hopf}, or by explicit
considerations of divergent sectors \cite{habil}, or
differential equations on bare Green functions \cite{CK}.


At such a level, one obtains a resolution of overlapping divergent
graphs into a sum of rooted trees, to which then the combinatorics
of the Hopf algebra of rooted trees applies \cite{hopf,CK}.


It was suggested to construct a Hopf algebra which directly
considers overlapping divergent graphs, without using
external input
as Schwinger Dyson equations \cite{KW}.
However, as already mentioned in \cite{CK},
this leads to the same Hopf algebra as for the case of
non-overlapping divergences, as we will prove
by  set-theoretic considerations.
\section{The Hopf algebra ${\cal H}_R$}
In this section we first repeat the definition of the Hopf
algebra of decorated rooted trees, as it can be found in \cite{CK}.
The rooted trees provide sets of vertices connected by edges.
The vertices are labelled by decorations.


Each decoration corresponds to an analytic expression
with a non-vanishing superficial degree of divergence, but 
free of subdivergences. Such analytic expressions are typically
obtained from general Feynman graphs by shrinking superficially
divergent subgraphs to a point. If for example
$\Gamma$ is a superficially divergent 
Feynman graph which contains only one divergent subgraph $\gamma$,
then one usually denotes by $\Gamma/\gamma$ the expression
in which $\gamma$ is reduced to a point in $\Gamma$.
When we speak of Feynman graphs in the following, this includes
such quotients $\Gamma/\gamma$.



The Hopf algebra of decorated rooted trees, 
with vertices labelled by Feynman graphs
free of subdivergences, is equivalent to the Hopf
algebra on parenthesized words introduced in
\cite{hopf}. In the next section, we 
embark on some set-theoretic considerations, which will
prove useful in the study of overlapping divergences.
In particular, we will assign a unique rooted tree
to a set $M$ by imposing conditions on its
subsets.


We follow section II of \cite{CK}.
A 
{\em rooted tree} $t$ is a connected and simply-connected set of oriented edges and 
vertices such that there is precisely one 
distinguished vertex which has  no incoming 
edge. 
This vertex is called the root of $t$. 
Further, every edge connects two vertices and
the {\em fertility} $f(v)$ of a vertex $v$ is the number of edges
outgoing from $v$. The trees being simply-connected, each vertex apart
from the root has a single incoming edge.


As in \cite{CK},
we consider the (commutative) algebra of polynomials over ${\bf Q}$ in rooted trees,
hence the multiplication $m(t,t^\prime)$ of two rooted trees means drawing them next to
each other in arbitrary order.


Note that for any rooted tree $t$ with root $r$ 
we have $f(r)$ trees $t_1$, $\ldots$,
$t_{f(r)}$ which are the trees attached to $r$.
The unit element of this algebra is $1$, corresponding,
as a rooted tree, to the empty set.



Let $B_-$ be the operator which removes the root
$r$ from a tree $t$:
\begin{equation}
B_-: t\to B_-(t)=t_1 t_2\ldots t_{f(r)}.
\end{equation}
Fig.(\ref{guill-}) gives an example.
\bookfig{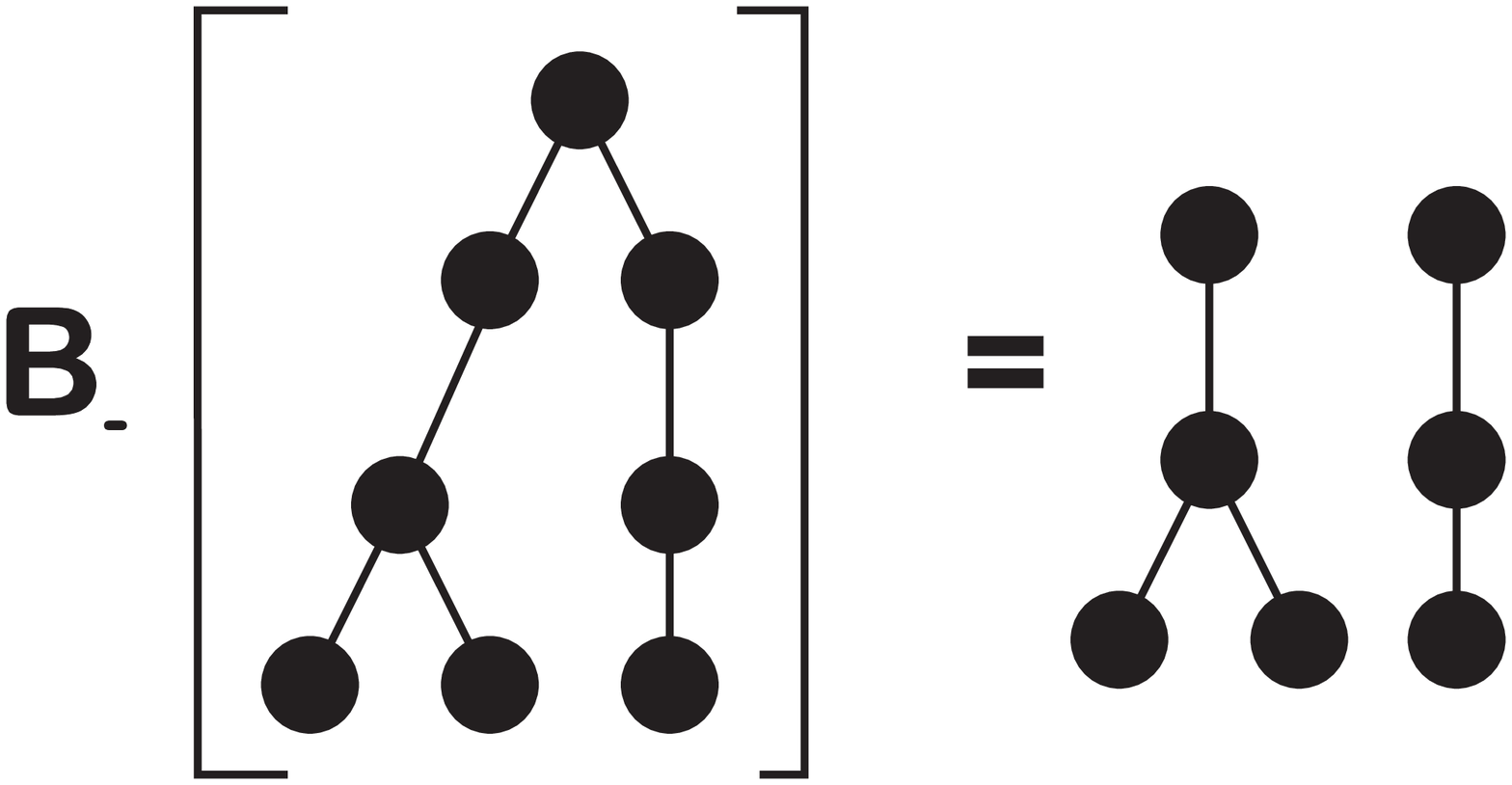}{$B_-$}{guill-}{The action
of $B_-$ on a rooted tree.}{3}


Let $B_+$ the operation which maps a monomial of $n$ rooted trees to a new rooted
tree $t$ which has a root $r$ with fertility $f(r)=n$ which connects to the $n$ 
roots 
of $t_1,\ldots,t_n$.
\begin{equation}
B_+: t_1\ldots t_n\to B_+(t_1\ldots t_n)=t.
\end{equation}
This is clearly the inverse to the action of $B_-$.


\bookfig{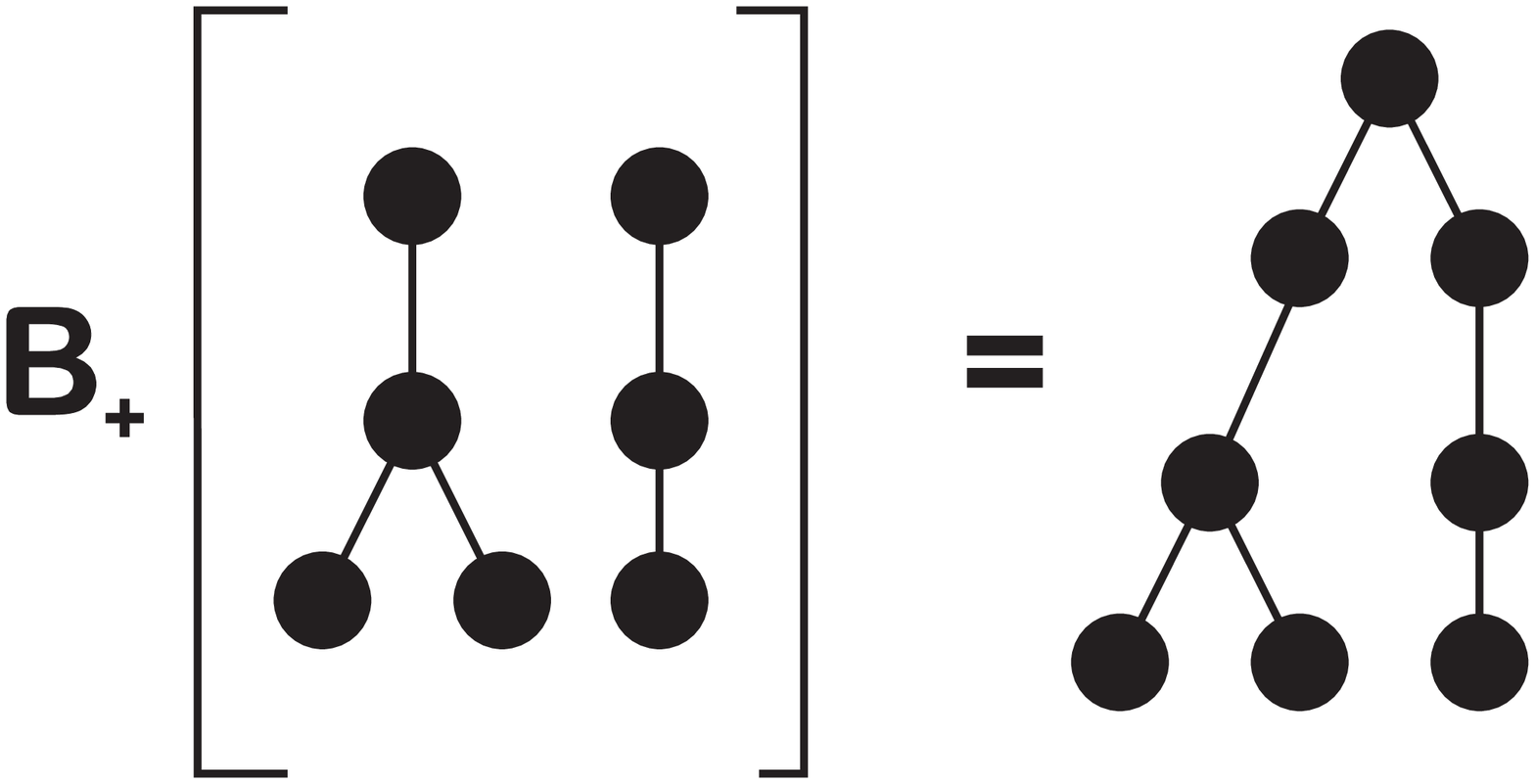}{$B_+$}{guill+}{The action
of $B_+$ on a monomial of  trees.}{3}


One has
\begin{equation}
B_+(B_-(t))=B_-(B_+(t))=t
\end{equation}
for any rooted tree $t$. Fig.(\ref{guill+}) gives an example.
For convenience, we define the rooted trees
$t_1,t_2,t_{3_1},t_{3_2}$ to be the trees with one, two or three vertices,
given in Fig.(5) on the lhs from top to bottom.


We further set $B_-(t_1)=1$, $B_+(1)=t_1$.


We will introduce a Hopf algebra on such rooted trees by using the possibility
to cut such trees in pieces. We start with the most elementary possibility.
An {\em elementary cut} is a cut of a rooted tree at a single chosen edge, as 
indicated in Fig.(\ref{ecut}). 
By such a cutting procedure, we will obtain the possibility to
define a coproduct, 
as we can use the resulting pieces on either side 
of 
the coproduct.


\bookfig{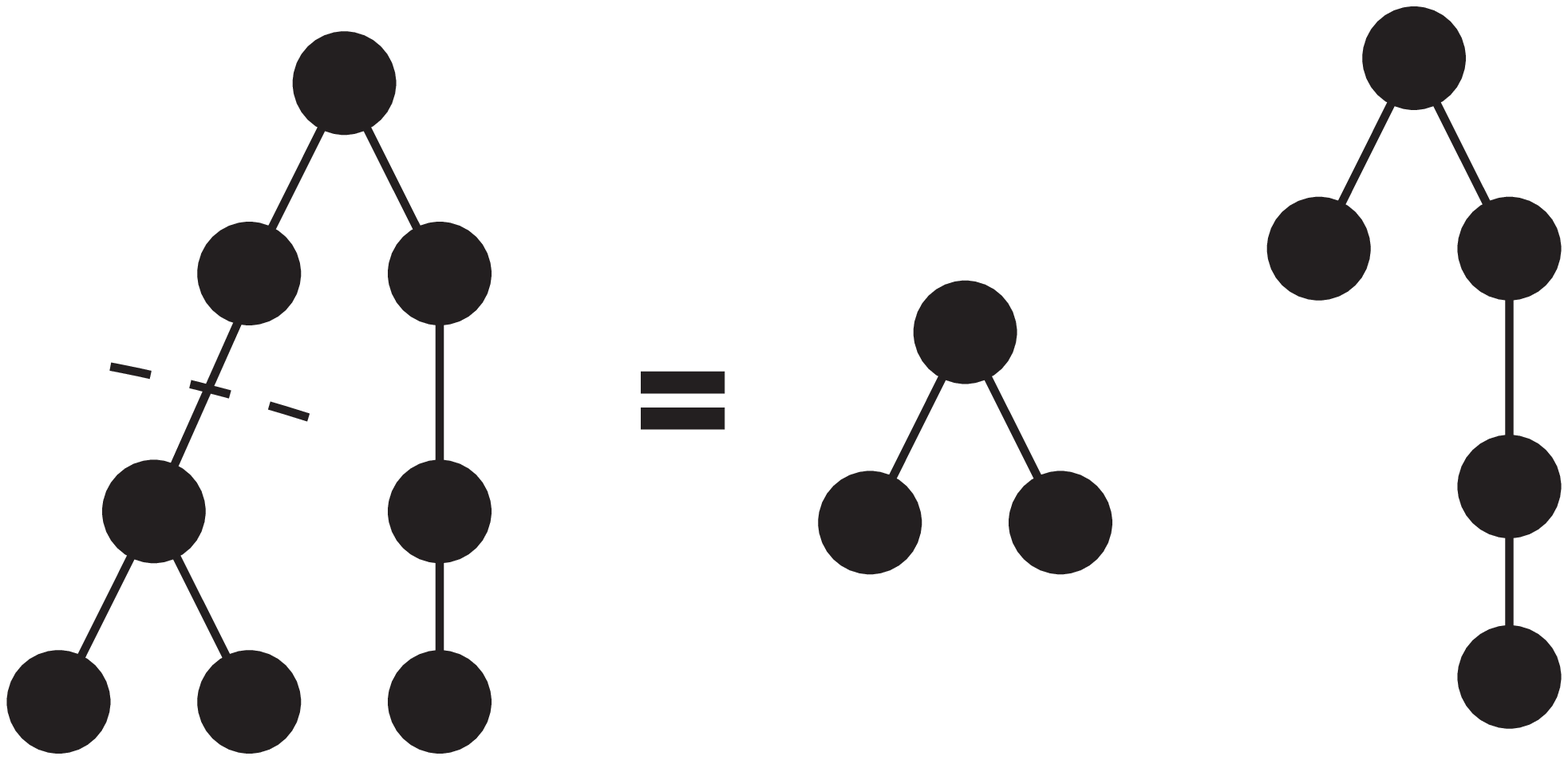}{Elementary cut.}{ecut}{An elementary cut $c$ splits a rooted
tree $t$ into two components, the fall-down $P^c(t)$ and the
piece which is still connected to the root, $R^c(t)$.}{3}


But before doing so we  finally introduce the notion of
an {\em admissible cut}, also called a {\em simple cut}. 
It is any assignment of elementary cuts to a rooted tree $t$ such that any path from
any vertex of the tree to the root has at most one elementary cut. Fig.(\ref{cut}) 
gives an example.


\bookfig{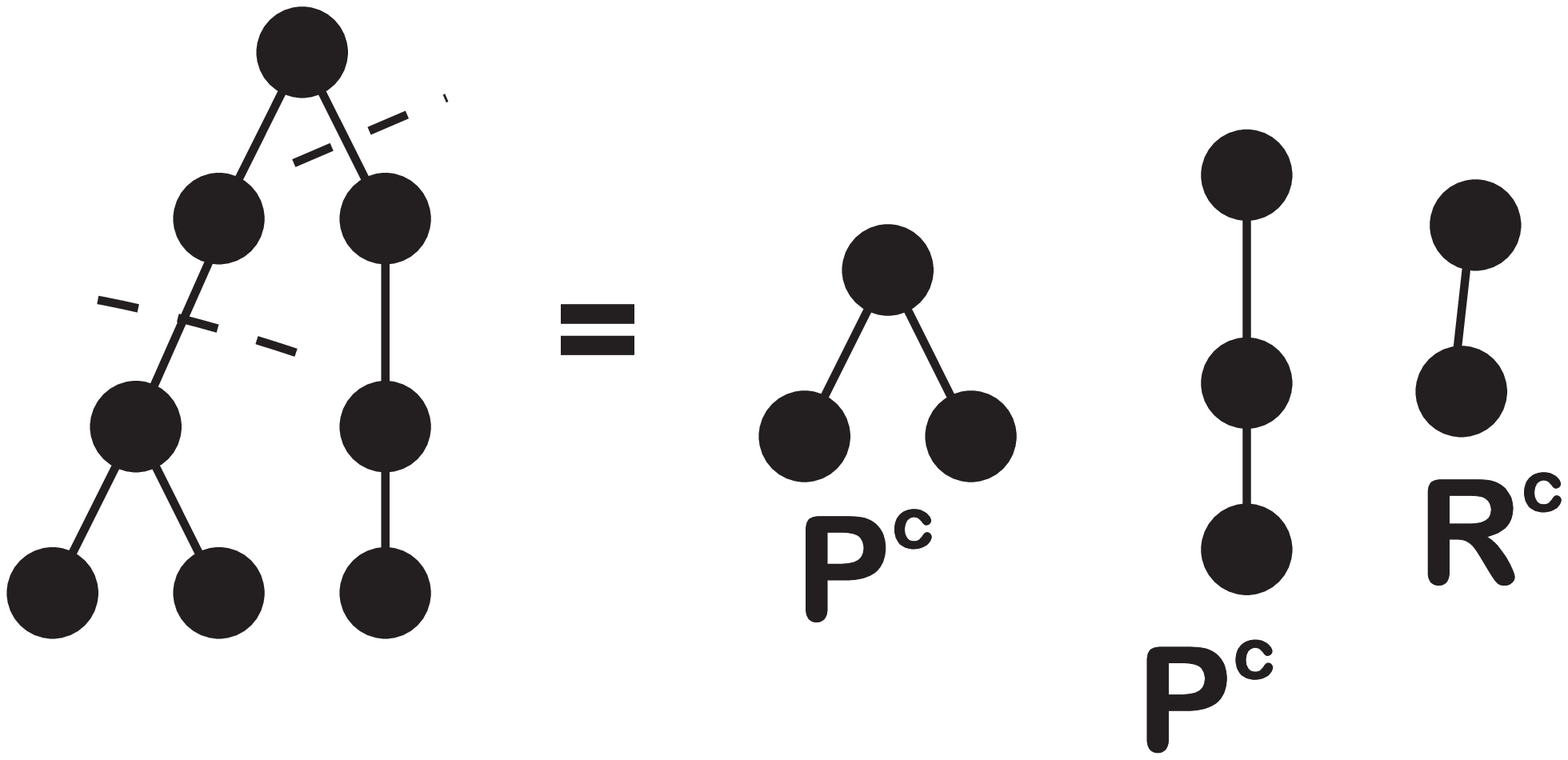}{An admissible cut.}{cut}{An
admissible cut $C$ acting
on a tree $t$. It produces a monomial of trees.
One of the factors, $R^C(t)$, contains the root of $t$.}{4}


An admissible cut $C$ maps a tree to a monomial in trees.
If the cut $C$ contains $n$ elementary cuts,
it induces a map
\begin{equation}
C: t\to C(t)=\prod_{i=1}^{n+1} t_{j_i}.
\end{equation}


Note that precisely one of these trees $t_{j_i}$
will contain the root of $t$.
Let us denote this distinguished tree
by $R^C(t)$. The monomial which is delivered by the
$n-1$ other factors is denoted by $P^C(t)$.


The definitions of $C,P,R$ can be extended to monomials of trees in the obvious 
manner, by choosing a cut $C^i$ for every tree $t_{j_i}$ in the monomial:
\begin{eqnarray*}
C(t_{j_1}\ldots t_{j_n}) & := & C^1(t_{j_1})\ldots C^n(t_{j_n}),\\
P^C(t_{j_1}\ldots t_{j_n}) & := & P^{C^1}(t_{j_1})\ldots
P^{C^n}(t_{j_n}),\\
R^C(t_{j_1}\ldots t_{j_n}) & := & R^{C^1}(t_{j_1})\ldots
R^{C^n}(t_{j_n}).
\end{eqnarray*}




Let us now establish the Hopf algebra structure. Following
\cite{hopf,CK} 
we define the 
counit and the coproduct. The {\em counit} $\bar{e}$: ${\cal A} \to {\bf Q}$ is 
simple:
$$
\bar{e}(X)=0
$$
for any $X\not= 1$,
$$
\bar{e}(1)=1.
$$


The {\em coproduct} $\Delta$
is defined by the equations
\begin{eqnarray}
\Delta(1) & = & 1\otimes 1\\
\Delta(t_1\ldots t_n) & = & \Delta(t_1)\ldots 
\Delta(t_n)\\
\Delta(t) & = & t \otimes 1 +(id\otimes B_+)[\Delta(B_-(t))],\label{cop2}
\end{eqnarray}
which defines the coproduct on trees with $n$ vertices
iteratively through the coproduct on trees with a lesser number of vertices.




\bookfig{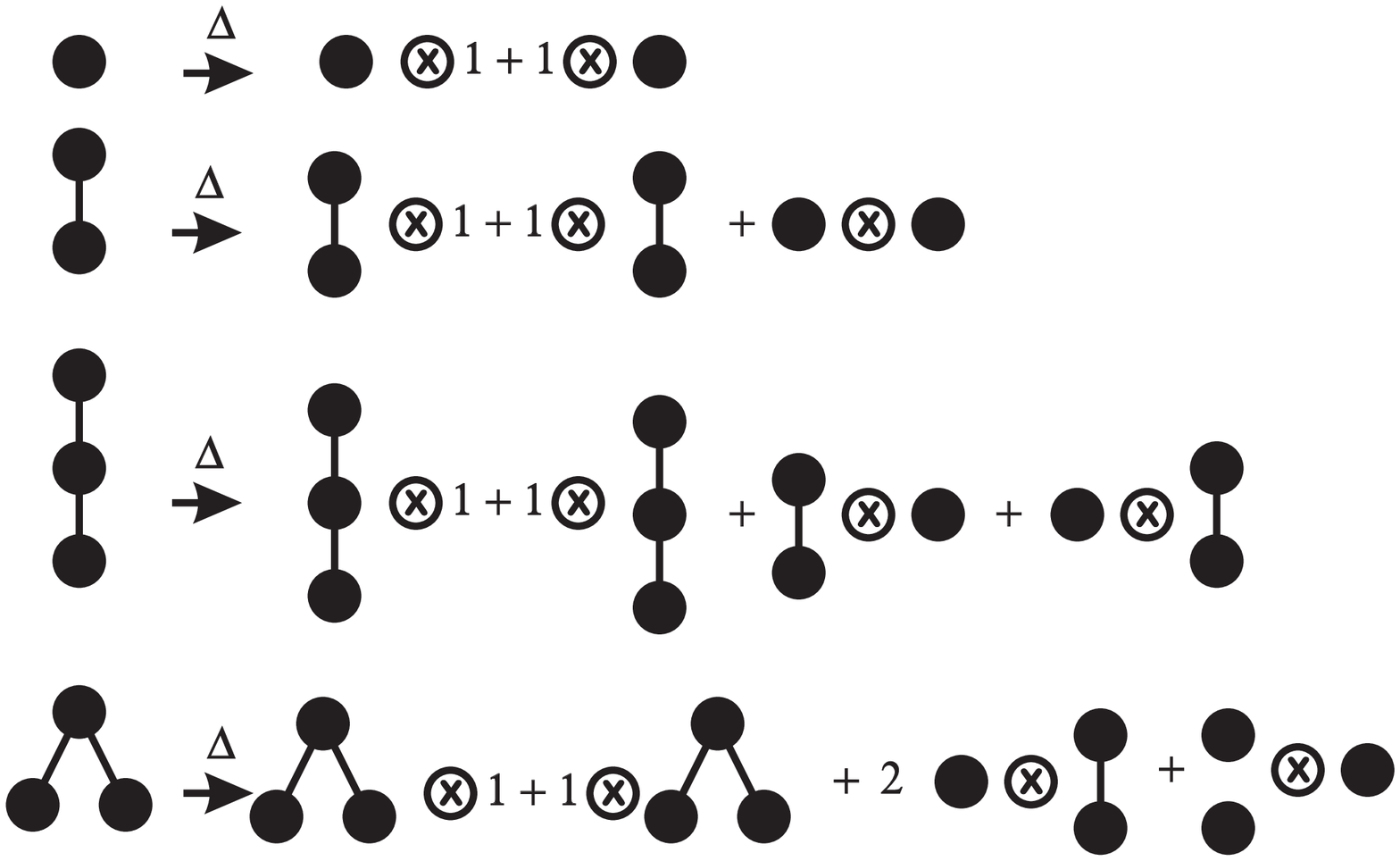}{The coproduct.}{cop}{The
coproduct. We work it out for the trees
$t_1,t_2,t_{3_1},t_{3_2}$, from top to bottom.}{6}


The coproduct can be written as \cite{hopf,CK}
\begin{equation}
\Delta(t)=1\otimes t+ t\otimes 1+
\sum_{\mbox{\tiny adm.~cuts $C$ of $t$}}P^{C}(t)\otimes R^C(t).\label{cop1}
\end{equation}






Up to now we have established a bialgebra structure. It is actually a Hopf algebra.
Following \cite{hopf,CK} we find the antipode $S$ as
\begin{eqnarray}
S(1) & = & 1\\
S(t) & = & -t-\sum_{\mbox{\tiny adm.~cuts $C$ of $t$}}S[P^C(t)]R^C(t).
\end{eqnarray}




Let us give yet another formula to write the antipode, which one easily derives 
using 
induction on the number of vertices \cite{hopf,CK}:
\begin{eqnarray*}
S(t) & = & -\sum_{\mbox{\tiny all  cuts $C$ of $t$}}(-1)^{n_C}P^C(t)R^C(t),
\end{eqnarray*}
where $n_C$ is the number of single cuts in $C$.


This time, we have a non-recursive expression, summing over all  cuts $C$, 
relaxing the restriction to admissible cuts.



By now we have established a Hopf algebra on rooted trees,
using the set of rooted trees, the commutative multiplication
$m$ for elements of this set, the unit $1$ and counit
$\bar{e}$, the coproduct $\Delta$ and antipode $S$.
We call this Hopf algebra ${\cal H}_R$.
Still following \cite{hopf,CK}
we allow to label the vertices of rooted trees by Feynman graphs without
subdivergences, in the sense described
before. 


Quite general, if $Y$ is a set of primitive elements
providing labels, 
we call the resulting Hopf algebra ${\cal H}_R(Y)$.
Let us also mention that
\begin{equation}
m[(S\otimes id)\Delta(t)]=\bar{e}(t)=0=\sum S(t_{(1)})t_{(2)},
\end{equation}
where we introduced Sweedler's notation
$\Delta(t)=:\sum t_{(1)}\otimes t_{(2)}$,
and $id$ is the identity map ${\cal H}_R\to{\cal H}_R$.



We finally note the following definition: for a rooted tree $t$
let $n_v(t)$ be the number of its vertices.
This extends to a monomial of rooted trees in the obvious manner,
$n_v(\prod_i T_i)=\sum_i n_v(T_i)$.


Ultimately, we work in the vector space of finite linear combinations
of monomials in rooted trees.
Hence for such a linear combination  $T:=\sum_i q^i X_i, i\in {\bf I}$,
for some index set ${\bf I}$,
we define 
\begin{equation}
n_v(T):=max\{n_v(X_i)\mid i\in {\bf I}\}.\label{477}
\end{equation}


\section{A set theoretic approach}
\subsection{Notation}
Let $\#(M)$ be the cardinality of any set $M$.\\
For any given finite set $M$ we let ${\cal P}(M)$ be the set of all 
proper subsets
of $M$. Further, we let 
\[
{\cal P}_X(M)\subset {\cal P}(M)
\]
be the set of all proper subsets of $M$ which fulfill
the condition $X$. Thus, 
if ${\bf X}$ is the boolean operator which is true when the condition
$X$ is satisfied, we have
\[
{\cal P}_X(M)=\{\gamma\in {\cal P}(M)\mid {\bf X}(\gamma)\}.
\]
If we impose no condition we write  $X=\emptyset$, hence  
${\cal P}_\emptyset(M)
\equiv{\cal P}(M)$.
If we want to stress that a subset $\gamma\subset M$
fulfills condition $X$, we write $\gamma\subsetX M$.



Let $\gamma_i,\gamma_j\subset{\cal P}(M)$, $i\not= j$, be two elements of 
${\cal P}_X(M)$, hence two subsets of $M$.
If 
\[\gamma_i\cap\gamma_j=\emptyset
\]
we call $\gamma_i,\gamma_j$
disjoint.\\
Else, if 
\[
\gamma_i\subset\gamma_j \;\mbox{or}\; \gamma_j\subset\gamma_i
\]
we call $\gamma_i,\gamma_j$
nested.\\
Finally, if $\gamma_i,\gamma_j$ are neither disjoint nor nested, we call them
overlapping. They then have a nontrivial intersection
$U:=\gamma_i\cap\gamma_j\not=\emptyset$, which is a
proper subset of each, $\gamma_i\supset U \subset \gamma_j$.\\
If $\gamma_i,\gamma_j$ are not overlapping, we call them
tree-related, for reasons which become obvious in a moment.\\
For a given set $X$ of mutually tree-related sets $\gamma_i$,
we say that another set $\gamma$ is overlapping with $X$ if $\gamma$
is overlapping with at least one element of $X$.



If a set $\gamma\subset {\cal P}_X(M)$ 
can be written as a union of {\em mutually disjoint} sets $\gamma_i\subset 
{\cal P}_X(M)$,
\[
\gamma=\cup_{i\in {\bf I}}\;\gamma_i
\] 
for some index set ${\bf I}$,
we say that $\gamma$ is reducible. Otherwise, we say it is
irreducible (w.r.t.$X$). 
Note that reducibility depends on the chosen
condition $X$.


Let $M/\gamma$ denote the complement of the set $\gamma\subset M$ with respect to
$M$, 
\[
M=M/\gamma\cup\gamma.
\] 
\subsection{Basic Results}
It is our task to find all  elements $p$ $\in$ 
${\cal P}({\cal P}_X(M))$
which  fulfill the following three conditions\\
i) $p$ consists of mutually tree-related sets $\in$ ${\cal
  P}_X(M)$,\\
ii) all elements of $p$ are irreducible,\\
iii) $p$ is complete: for all $\gamma\subset {\cal P}_X(M)$
such that $\gamma\not\in p$ $\Leftrightarrow$
$\gamma$ is overlapping with $p$. 


For an irreducible $M$,
let the set of all such $p$, that is the set of all
{\em c}omplete, {\em i}rreducible, {\em t}ree-ordered elements of
${\cal P}({\cal P}_X(M))$ be denoted by 
${\cal P}_X^{cit}(M)$.


\smallskip


\noindent {\bf Prop.1}
To each such $p\in {\cal P}_X^{cit}(M)$, 
we can assign a rooted tree $T_X(p)$ with $n=(\#(p)+1)$ vertices.\\
{\bf Proof:}
We draw $n$ points in the plane, which furnish the set of
vertices
of the rooted tree.
To one of these points, 
we associate the set $M$. It will become the root.
To each of the  other $n-1$ points we associate one element of $p$.
Let $v(\gamma_i)$ denote the vertex which is labelled by the
set $\gamma_i\in p$ in this process.


Now we can  construct the edges. 
For that, we connect two vertices $v(\gamma_i),v(\gamma_k)$
by an edge pointing from
$v(\gamma_k)$ to $v(\gamma_i)$
if and only if the following two conditions are fulfilled:\\
i) $\gamma_i\subset
\gamma_k$,\\
ii) there is no further set $\gamma_j\in p$
such that $\gamma_i\subset\gamma_j\subset \gamma_k$.\\ 
Here, we allow $\gamma_k$ to be the set $M$ itself:
 $\gamma_k\in \{p\cup M\}$.\\
The resulting tree is simply-connected,
due to the fact that all elements of $p$ are mutually tree-ordered.
Further, it has a distinguished root. ~ $\Box$


\smallskip



For a chosen vertex $v$ of a rooted tree $T_X(p)$
let $\gamma(v)$ be the set associated to that vertex.
Further, assume that $f(v)=k$, hence
$v$  connects via $k$ outgoing edges
to vertices $v_1,\ldots,v_k$, say.
The corresponding sets $\gamma(v_1),\ldots,\gamma(v_k)$
are necessarily mutually disjoint, as $T_X(p)$ is simply-connected.


Define 
\[
\gamma_v:=\cup_{i:=1}^{f(v)} \gamma(v_i).
\]


\smallskip


\noindent {\bf Prop.2} 
${\cal P}_X(\gamma(v)/\gamma_v)$ $=$ $\emptyset$.\\
{\bf Proof:} 
$\gamma(v)$ is irreducible as it is an element of $p$.
Hence $\gamma(v)/\gamma_v\not=\emptyset.$ 
By definition, $\gamma_v$ is the union of all sets $\gamma(v_i)$ $\in$
$p$
which are subsets $\gamma(v_i)\subset \gamma(v)$.
If there would be an element $\gamma^\prime$  in
${\cal P}_X(\gamma(v)/\gamma_v)$, this would imply
that $\gamma^\prime$ is a non-overlapping subset 
of $\gamma(v)$ which is not in $\gamma_v$. Contradiction.
 ~
$\Box$


\smallskip 


The linear combination   $T_X(M)$ assigned to the irreducible set $M$
is the sum 
\[
T_X(M):=\sum_{p\in {\cal P}_X^{cit}(M)}T_X(p).
\] 



For a reducible $M$ we can write $M=\cup_i M_i$ for some
mutually disjoint irreducible sets $M_i$.
We then set $T_X(M)=\sum_i T_X(M_i)$. 
\footnote{If there is
more than one possibility to write $M$ as a union
of disjoint sets $M_i$, we sum over all trees which we obtain
from the consideration of
all possibilities how to decompose $M$ into these various
disjoint subsets. We will not meet this case in this paper, though.}




An example might be in order.
Let $M=\{a,b,c\}$. First, choose $X=\emptyset$.
All subsets which contain more than one element are reducible.
Thus, $T_X(M)$ is the product  $t_1(a)t_1(b)t_1(c)$
of three disjoint roots,
labelled $\{a\},\{b\},\{c\}$.


Next, let $X$ be the condition that 
$a$ is contained in the subset but not $c$.
Then, $\{a\}$ and $\{a,b\}$ are irreducible proper subsets.
${\cal P}_X^{cit}(M)$ contains a single set
$p=\{\{a\},\{a,b\}\}$, and we obtain $T_X(p)=t_{3_1}$ (see Fig.(5))
with the set $M$ labelling the root, which is connected
 to a vertex labelled by $\{a,b\}$, and finally this vertex is
 connected
to a third one labelled by
$\{a\}$. 


Finally, choose $X$ to be the condition that
$a$ is contained in the subset.
Then, $\{a\},\{a,b\},\{a,c\}$ are irreducible proper
subsets. The latter two are overlapping.
${\cal P}_X^{cit}(M)$ consists of two elements $p_1,p_2$,
say, where
$p_1=\{\{a\},\{a,b\}\}$ and $p_2=\{\{a\},\{a,c\}\}$.
$T_X(p_1)$ and $T_X(p_2)$ both realize $t_{3_1}$
with appropriate decorations.
Consider Fig.(\ref{sets}) for a visualization of these
examples.
\bookfig{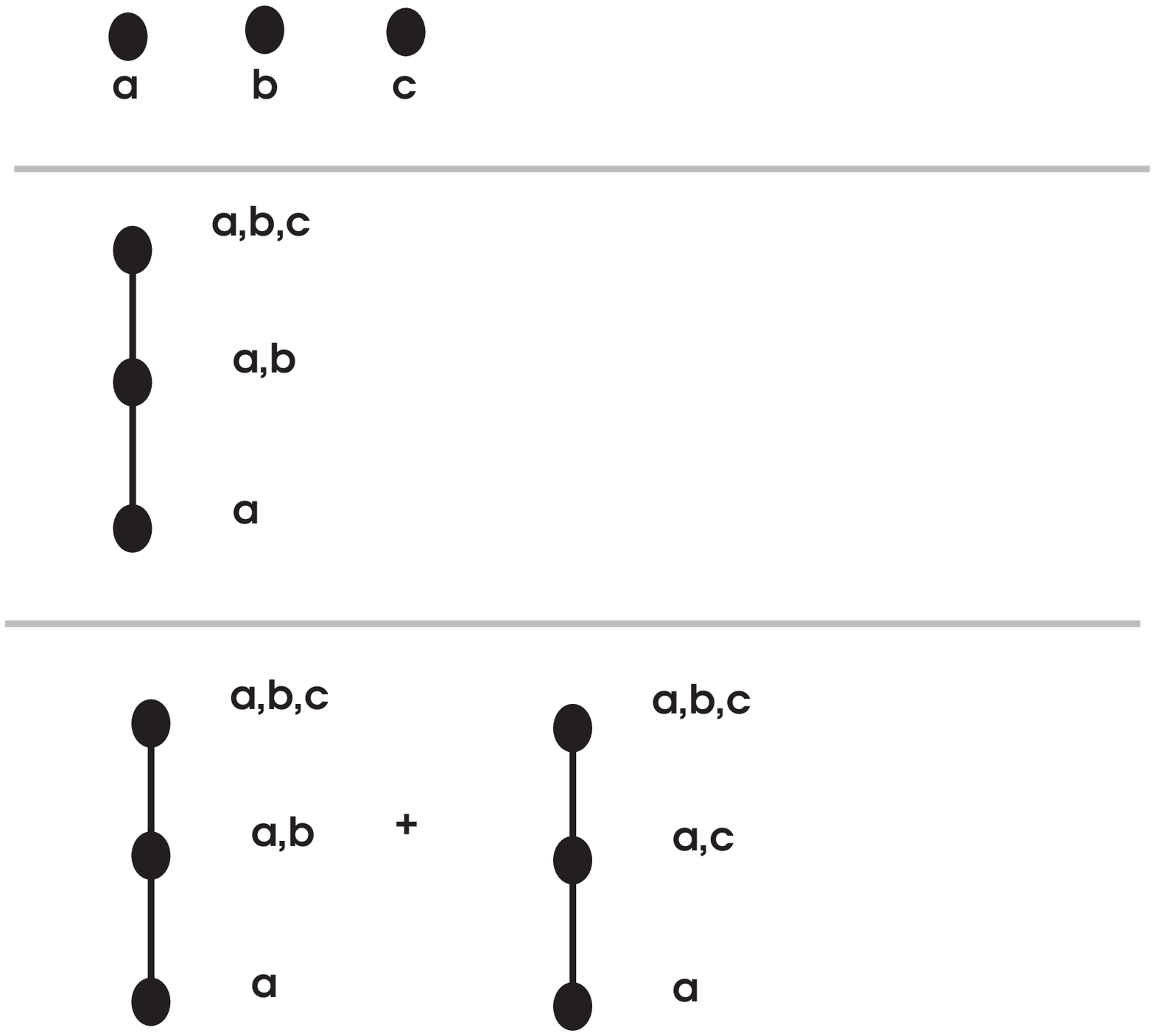}{Examples.}{sets}{The examples show how various $T_X(\{a,b,c\})$
are generated by different conditions $X$.
From top to bottom, we have i)$X=\emptyset$, ii)$X$: $a\in \gamma,
c\not\in\gamma$, iii)$X: a\in\gamma$.}{5}
\subsection{The Hopf algebra structure of ${\cal P}_X(M)$}
To each set $M$ we can assign a depth $d_X(M)$ as
 $d_X(M)=n_v(T_X(M))$, according to (\ref{477}).
This gives us a decomposition on the set ${\cal M}$ of all 
irreducible (w.r.t.$X$) finite sets through the grading by depth,
\[
{\cal M}={\cal M}^{[1]}\cup{\cal M}^{[2]}\cup\ldots\cup{\cal M}{[k]}\cup\ldots,
\]
which obviously depends on the condition $X$.
Here, ${\cal M}^{[1]}$ are all sets $M$ which have no proper subset
which fulfills condition $X$, hence of depth $d_X(M)=1$,
${\cal M}^{[2]}$ are all sets of depth two,
such that all their proper subsets which fulfill
$X$ are from ${\cal M}^{[1]}$, and in general
${\cal M}^{[k]}$ contains all sets of depth $k$, and hence has proper
subsets of depth $\leq$ $k-1$.


We want to establish a Hopf algebra of rooted trees on ${\cal M}$.
For this, it is sufficient to study irreducible sets $M$.
We will take elements of ${\cal M}^{[1]}$ as primitive elements.
By definition, $T(M)=t_1(M)$ for $M\in {\cal M}^{[1]}$, which justifies
this choice.


We call a set $M$ non-overlapping, if ${\cal P}_X(M)$
is tree-ordered,
hence if all its subsets which fulfill $X$ are tree related amongst each other.


\smallskip 


\noindent {\bf Prop.3} If $M$ is non-overlapping,
$\#({\cal P}_X^{cit}(M))=1$.\\
{\bf Proof:} All elements of ${\cal P}_X(M)$
can be tree-ordered amongst themselves by assumption.
As any element $p\in{\cal P}_X^{cit}(M)$ is complete
and contained in ${\cal P}_X(M)$, there can be only one such
element $p$. ~ $\Box$


\smallskip


\noindent Two final definitions: If $X$ is a given condition,
\[
{\cal P}_X(M)=\{u\in {\cal P}(M)\mid {\bf X}(u)\},
\]
then, for $\gamma\in {\cal P}_X(M)$,
$X_\gamma$ is defined to be the condition
\[
{\cal P}_{X_\gamma}(M)=
\{u\in {\cal P}(M)\mid {\bf X}(u)\; \mbox{and}\;u\not\in 
\{\gamma\cup{\cal P}_X(\gamma)\}\; \}.
\]
We call a condition $X$ an orderly condition if and only if 
$T_{X_\gamma}(M)=T_X(M/\gamma)$, $\forall \gamma\in{\cal P}_X(M)$.
This means that checking the condition $X$ and then eliminating all
elements of ${\cal P}_X(M)$ which belong
as well to $\gamma\cup{\cal P}_X(\gamma)$ is the same
as first
eliminating $\gamma$ and
checking the condition $X$ on the reduced set $M/\gamma$.



Let us give an example of an orderly condition.


\smallskip


\noindent {\bf Example:}
Consider a space $Y$ and a set $\sigma_Y$ of subsets of $Y$.
Endow $Y$ with the topology generated by $\sigma_Y$
as a subbasis.\footnote{Any set
$\sigma$ 
of subsets of a space generates a topology. The open sets are
unions of finitely many intersections of elements of $\sigma$,
and $\sigma$ is the subbasis of this topology.}
Endow any space $Y/\gamma$, $\gamma\in\sigma_Y$ with its induced
topology, which is generated by the subbasis
$\{u/\gamma\mid u\in\sigma_Y\}$. Let $X$ be the condition 
that a subset $\gamma\subset Y$ must fulfill $\gamma\in
\sigma_Y$ to be in ${\cal P}_X(Y)$. Then,
$X$ is an orderly condition.
Indeed, $T_X(Y/\gamma)$ is the forest $T_X(Y)$ in which
all vertices decorated by $\gamma$
or its subsets in ${\cal P}_X(\gamma)$
are deleted, and so is $T_{X_\gamma}(Y)$.


\smallskip



\noindent On the other hand, note that the examples in Fig.(\ref{sets})
give non-orderly conditions for $X\not=\emptyset$.



We want to establish a Hopf algebra of rooted
trees ${\cal H}_R({\cal M}^{[1]}\cup_{i=2}^\infty U^{[i]})$
which assigns
to each $M\in {\cal M}^{[k]}$
a sum of rooted trees $T_M$ such that its coproduct takes the form
\begin{equation}
\Delta(T_M)=\sum_{\gamma\subsetX M}T_\gamma\otimes T_{M/\gamma}.\label{natcop}
\end{equation}
The sum is over all subsets $\gamma\subset M$ such that  
$\gamma$ fulfills
condition $X$.
We do not demand that $\gamma$ is irreducible.
It is thus allowed that $\gamma$ 
 is the union of disjoint sets
$\gamma_i$  which  themselves fulfill condition $X$ and are
 irreducible.


The notation $\cup_{i=2}^\infty U^{[k]}$ refers to the iterative manner in which we will
achieve our goal. To achieve our goal
for sets $M$ of depth one is trivial.
We take ${\cal M}^{[1]}$ as the set of decorations
for ${\cal H}_R$ and are done.
Next, we will construct a set of decorations,
$U^{[2]}$ such that ${\cal H}_R({\cal M}^{[1]}\cup U^{[2]})$
achieves the desired goal for all sets of depth up to two.
Then, we further enlarge this set by $U^{[3]}$ so that
the coproduct in ${\cal H}_R({\cal M}^{[1]}\cup U^{[2]}\cup U^{[3]})$
agrees with (\ref{natcop}) for sets $M$ of depth up to three
and so on. In general, we show that if one has succeeded
at depth $k$ that there is a set of decorations $U^{[k+1]}$
which are primitive under the coproduct of ${\cal H}_R$,
such that one obtains the desired form (\ref{natcop}).


It will turn out that $T_M$ is a sum of rooted trees
containing $T_X(M)$. Further, $[T_M-T_X(M)]$ 
is a sum of rooted trees
which fulfills $n_v(T_M-T_X(M))<n_v(T_X(M))$.




For non-overlapping sets $M$, there is an immediate 
natural Hopf algebra structure
${\cal H}_R({\cal M}^{[1]})$.
It is natural in the sense that the coproduct assumes the form
(\ref{natcop}):


\noindent {\bf Prop.4} For non-overlapping sets $M$
we have 
\begin{equation}
\Delta(T_X(M))
=\sum_{\gamma\subsetX M}T_X(\gamma)\otimes T_{X_\gamma}(M).\label{natcopg}
\end{equation}
{\bf Proof:} For non-overlapping sets $M$, $T_X(M)$ is a single
rooted tree $T_X(M)=T_X(p)$.
Admissible cuts on this rooted tree and subsets $\gamma$ in the sum are
in one-to-one correspondence, by construction.
Let $\gamma_C$ be the set corresponding
to the chosen admissible cut $C$.
By the definition of $T_X(\gamma)$, $T_X(\gamma)=P^C(T_X(M))$.
Further, $R^C(T_X(M))$ is the decorated tree which remains 
connected with the root under the admissible cut.
By definition of $X_\gamma$,  $T_{X_\gamma}(M)=R^C(T_X(M))$
as both rooted trees are obtained from $T_X(M)$ 
by eliminating all vertices
and edges corresponding to $T_X(\gamma)$.
Further, by 
Prop.2 we can decorate the rooted tree $T_X(M)$ with elements from
${\cal M}^{[1]}$. ~ $\Box$


\smallskip


Note that for an orderly condition $X$, (\ref{natcopg}) takes
the form
\begin{equation}
\Delta(T_X(M))=\sum_{\gamma\subsetX M}T_X(\gamma)\otimes T_X(M/\gamma).
\label{natcopn}
\end{equation}
Hence we set 
\begin{equation}
T_\gamma=T_X(\gamma),\;
T_{M/\gamma}=T_X(M/\gamma),\label{set}
\end{equation} to obtain the desired form
(\ref{natcop})
for all non-overlapping $M\in {\cal M}^{nol}$,
the set of all sets $M$ which are non-overlapping. 
This is consistent as if $M$ is non-overlapping,
so are all elements in ${\cal P}_X(M)$.


To simplify notation, 
let us assume in the following that $X$ is an orderly condition.
When we come to Feynman graphs in the next section, 
we will actually find  the relevant condition
$X$ to be
an orderly condition. 
However, the general case demands not much more
than a replacement $T_X(M/\gamma)\to T_{X_\gamma}(M)$
and a slightly refined decomposition of ${\cal M}$.


So far, we found that all elements in ${\cal M}^{nol}$
have the desired form. From now on let $\Delta_1$
be the coproduct
of ${\cal H}_R({\cal M}^{[1]})$. We have just shown that it has the
desired
form on ${\cal M}^{nol}$. We stress that $\Delta_1$ is defined on
all rooted trees with decorations in ${\cal M}^{[1]}$.


We now want to show that for the other elements,
which are overlapping sets $M$, we can find a 
Hopf algebra of rooted trees with a coproduct
which has the desired form (\ref{natcop}), by simply adding more
decorations. 
As an aside, we will gain
a systematic decomposition into primitive elements, which corresponds
to a skeleton expansion at the level of QFT, as we will see later on.


We will proceed by induction on the depth. There are no overlapping
sets $M$ in ${\cal M}^{[1]}$, ${\cal M}^{[1]}\subset{\cal M}^{nol}$. 



Hence we start the induction
by considering sets in ${\cal M}^{[2]}$.
We want to construct a Hopf algebra of rooted trees ${\cal H}_R(
{\cal M}^{[1]}\cup U^{[2]})$ such that its coproduct $\Delta_2$
again
can be written in the form (\ref{natcop}).
$U^{[2]}$ is a set of decorations, hence we demand
$\Delta_2(u)=u\otimes 1+1\otimes u$,
$\forall u\in U^{[2]}$.


Let $M\in {\cal M}^{[2]}$ be irreducible and overlapping.
Then, each $p\in {\cal P}_X^{[cit]}(M)$ is in
${\cal M}^{[1]}$.


Let us assign to $M$ an element $T_M$ and set
\[
\Delta_2(T_M)=T_M\otimes 1+1\otimes T_M+
\sum_{\gamma\subsetX M} T_\gamma\otimes T_{M/\gamma}.
\]
Due to the definition of ${\cal P}_X^{cit}(M)$ 
this can be written as
\[
\Delta_2(T_M)=T_M\otimes 1+1\otimes T_M+
\sum_{p\in {\cal P}_X^{cit}(M)} T_p\otimes T_{M/p}.
\]
But $p\in {\cal M}^{[1]}$ and $M/p\in {\cal M}^{[1]}$,
hence $T_p=T_X(p)$, $T_{M/p}=T_X(M/p)$, by (\ref{set}).


Also, the coproduct $\Delta_1$
of ${\cal H}_R({\cal M}^{[1]})$ is defined
on the sum of rooted trees $T_X(M)$ and reads
\[
\Delta_1(T_X(M))=T_X(M)\otimes 1+1\otimes T_X(M)+
\sum_{p\in {\cal P}_X^{cit}(M)} T_X(p)\otimes T_X(M/p).
\]
Thus, we find that, for $U_M:=T_M-T_X(M)$,
\[
\Delta_2(U_M)=U_M\otimes 1+1\otimes U_M.
\]
$U_M$ reveals itself to be a primitive element with respect to $\Delta_2$.
This suggests to define $U^{[2]}$ via the union of all elements
$U_M=T_M-T_X(M)$ where $M$ is of depth two and overlapping.
As $U_M$ is primitive we identify it with a decoration $u_M$
of the tree $t_1$, $t_1(u_M)=U_M$ and obtain
\begin{equation}
U^{[2]}=\{u_M\mid t_1(u_M)=T_M-T_X(M),\;M\in {\cal M}^{[2]}\wedge M\not\in{\cal M}^{nol}\}
\label{u2}
\end{equation}
Hence we
find that  $\Delta_2$ is the coproduct of the Hopf algebra
of rooted trees
\[
{\cal H}_R({\cal M}^{[1]}\cup U^{[2]}),
\]
where $U^{[2]}$ is the set of decorations
corresponding to primitive
elements $U_M\equiv T_M-T_X(M)$, 
$M$ being an overlapping
set in ${\cal M}^{[2]}$.
The primitive elements of this Hopf algebra are 
$t_1(M)$, $M\in {\cal M}^{[1]}$ and the elements
$U_M$ defined above. 


Note that the element $T_M$ is resolved into the
linear combination of trees  $T_M=T_X(M)+t_1(u_M)$, as desired.
Note further that we can write the coproduct
$\Delta_2$ as
\[
\Delta_2(T_M)=T_M\otimes 1+1\otimes T_M+
(id-E\circ\bar{e})\otimes(id-E\circ\bar{e})\Delta_1(T_X(M)),
\]
where $E: {\bf Q}\to {\cal H}_R$ is given by $E(q)=q1$.
Obviously we left
the counit $\bar{e}$ unchanged, $\bar{e}(1)=1$,
$\bar{e}(T)=0$, $\forall T\not=1$.




At this point the attentive reader might ask why we not simply set
$T_M=T_X(M)$, as this would still deliver the natural form
(\ref{natcop}). But our point is to show that any attempt 
to find a Hopf algebra which has the natural form (\ref{natcop})
will be a Hopf algebra of rooted trees, with an appropriate set
of primitive elements. This completely puts the combinatorical
problem of renormalization at rest and settles its algebraic
structure as determined by the Hopf algebra structure of rooted trees,
which, fascinatingly, not only describes renormalization
but also the combinatorics of the diffeomorphism group \cite{CK}.


Let us continue then.
Thus, let $M\in{\cal M}^{[k]}$ be irreducible
and overlapping. 
Assume we found a Hopf algebra of rooted
trees 
${\cal H}_R({\cal M}^{[1]}\cup_{i=2}^k U^{[i]})$ with
coproduct $\Delta_k$ such that in this
Hopf algebra there is a linear combination 
$T_M$ of elements such that the coproduct obtains
the form (\ref{natcop}),
\[
\Delta_k(T_M)=T_M\otimes e+e\otimes T_M+\sum_{\gamma\subsetX M}
T_\gamma\otimes T_{M/\gamma}.
\]
We want to induce that the same holds for $M_+\in{\cal M}^{[k+1]}$.


Let $\gamma\subsetX M_+$ be given, and let
$M_+\in {\cal M}^{[k+1]}$ be overlapping.
Then, consider all the terms in 
\[
\Delta_1[T_X(M_+)]
=\sum_{p\in {\cal P}_X^{cit}(M_+)}\left[\sum_{\mbox{\tiny
adm.cts. $C^p$ of $T_X(p)$}}
P^{C(p)}[T_X(p)]\otimes R^{C(p)}[T_X(p)]\right],
\]
which correspond to the set $\gamma$. 
This is well-defined:  any two overlapping sets
$\gamma,\gamma^\prime\in{\cal P}_X(M_+)$ will correspond to branches of
different trees $T_X(p),T_X(p^\prime)$, as elements
$p,p^\prime$ are tree-ordered.  
Further, each single elementary
cut corresponds to some subset $\gamma\subsetX M_+$.
We can thus organize the above sum in groups of terms corresponding
to $\gamma\subsetX M_+$.
Finally, the completeness of all elements
of ${\cal P}_X^{cit}(M_+)$ 
guarantees that all
admissible cuts which correspond to $\gamma$ conspire to give
$T_X(\gamma)$, and all terms on the other side of the tensorproduct
conspire to give $T_X(M_+/\gamma)$, for an orderly condition $X$.
We get
\begin{eqnarray}
\Delta_1[T_X(M_+)] & = & \sum_{p\in {\cal P}_X^{cit}(M_+)}\left[\sum_{\mbox{\tiny
adm.cts. $C^p$ of $T_X(p)$}}
P^{C(p)}[T_X(p)]\otimes R^{C(p)}[T_X(p)]\right]\nonumber\\
 & = & \sum_{\gamma\subsetX M_+}T_X(\gamma)\otimes
T_X(M_+/\gamma).\label{copno}
\end{eqnarray}
Now we have to take care of the difference between $T_X(\gamma)$ 
and $T_\gamma$, and between $T_X(M_+/\gamma)$ and $T_{M_+/\gamma}$.


We first take care of all possible differences between
$T_X(\gamma)$ and $T_\gamma$. Consider all $\gamma\in{\cal P}_X(M_+)$.
First, we consider all such $\gamma$ which are in ${\cal M}^{[2]}$ and
overlapping. In the coproduct (\ref{copno}) we find a term
$T_X(\gamma)$ on the lhs. $T_X(\gamma)\otimes T_X(M_+/\gamma)$
is actually a sum of terms (as on both sides are sums of trees in
general) which carries a natural product structure indicated by the
tensorproduct. For each term in this sum, there is a well-defined
set of edges corresponding to the admissible cut which gives
$\gamma$. Gluing both sides,
$T_X(\gamma)$ and $T_X(M/\gamma)$, 
together along these edges gives back
$T_X(M_+)$,
$$
T_X(M)=T_X(\gamma)\wedge_\gamma T_X(M/\gamma).
$$ 
Here $\wedge_\gamma$ refers to the gluing process
along the edges which are cut when we obtain $T_X(\gamma)$
on the lhs of the tensorproduct.
Instead, we glue $T_\gamma=T_X(\gamma)+U_\gamma$
back along these edges ('surgery along edges'), 
for all overlapping $\gamma\in {\cal M}^{[2]}$.
Call the new sum of trees
$$
T_2(M_+)=T_\gamma\wedge_\gamma T_X(M/\gamma).
$$ 


It has the form $T_X(M_+)+T_2$,
where $n_v(T_2)=n_v(T_X(M_+))-1$. 


It further has the property that
all cuts corresponding to such a $\gamma$ in $T_2(M_+)$ will give
$T_\gamma$ on the lhs,
if we employ $\Delta_2[T_2(M_+)]$. 


Now we consider all $\gamma\in {\cal P}_X(M_+)$
which are overlapping and in ${\cal M}^{[3]}$.
We use the product structure of $T_2(M)$ under 
$\Delta_2$ and glue back  $T_{\gamma}$ for $T_2(\gamma)$.
We continue in this manner for all overlapping $\gamma\in{\cal P}_X(M_+)$
in ascending order until we reach $\gamma\in{\cal M}^{[k]}$.
Call the resulting sum of trees  $T_k(M)$. 


In a similar manner, we then replace $T_k(M/\gamma)$
by $T_{M/\gamma}$ starting with $M/\gamma\in {\cal M}^{[2]}$. 
We finally obtain a sum of trees $\breve{T}(M)$ $=T_X(M)+$  terms
of lower depth. 


By construction, $\Delta_k(\breve{T}(M)-T_X(M))$ 
contains all  the terms which distinguish
$\sum_{\gamma\subsetX M}T_\gamma\otimes T_{M/\gamma}$
from $\Delta(T_X(M))$. Notably, $\Delta_k$
acts on $\breve{T}(M)$, as it is a sum of rooted trees with 
decorations in ${\cal M}^{[1]}\cup_{i=2}^k U^{[i]}$.


Hence we get 
\[
\Delta_k(\breve{T}(M))=
\breve{T}(M)\otimes e +e \otimes 
\breve{T}(M)+\sum_{\gamma\subsetX M}T_\gamma\otimes T_{M/\gamma}.
\]
We now set
\[
\Delta_{k+1}(T_M):=T_M\otimes e+e\otimes T_M+
(id-\bar{e})\otimes (id-\bar{e})\Delta_k(\breve{T}(M)).
\]
Then, again, $U_M:=T_M-\breve{T}(M)$
is a primitive element for $\Delta_{k+1}$, and thus
$\Delta_{k+1}$ becomes the coproduct of a Hopf algebra of rooted trees
${\cal H}_R({\cal M}^{[1]}\cup_{i=2}^{k+1} U^{[i]})$,
where $U^{[k+1]}$ is the set of elements  $U_M$, with $M$  an overlapping
set of depth $k+1$, thus in ${\cal M}^{[k+1]}$.
We have, in analogy to the case $k=2$,
\begin{equation}
U^{[k+1]}=\{u_M\mid t_1(u_M)=T_M-\breve{T}(M),
\;M\in {\cal M}^{[k+1]}\wedge M\not\in{\cal M}^{nol}\},
\end{equation}
which works iteratively as $\breve{T}$ uses only decorations
obtained from trees with degree $\leq k$.


Hence, algorithmically, one needs
to determine all elements $\gamma\in {\cal P}_X(M)$,
and then all elements in the corresponding ${\cal P}_X(\gamma)$,
and so on. Eventually, one ends considering elements of depth two,
whose decorations can be immediately determined, by
(\ref{u2}). One then works up with the grading.



We conclude that the natural coproduct (\ref{natcop})
is the coproduct of the Hopf algebra of rooted trees
based on an appropriate set of decorations, constructed iteratively
starting at depth two and using induction on the grading by depth.


Regarding an arbitrary Feynman graph as a set of edges and vertices,
it needs powercounting and a determination of (one-particle
irreducible) subgraphs to determine 
the skeleton expansion,
and hence all decorations, iteratively, as the examples in the next
section will exhibit. 
\section{Overlapping Divergences}
We will apply the notions established in the previous section
to sets of propagators and vertices which constitute Feynman graphs.
We will order
Feynman graphs by the depth of the rooted
trees assigned to them. Below, we will define an orderly condition
$X$ which can be tested by powercounting.
The elementary fact that Feynman integrals allow for a well-defined
degree of divergence essentially allows to use this degree
of divergence as the crucial check on subgraphs, regarded as subsets
of edges and vertices. One-particle irreducibility
is the other demand which we choose for convenience.


For each one-particle irreducible (1PI)
superficially divergent Feynman graph
$\Gamma$ we denote by $\{\Gamma\}$ the set of its propagators and
vertices.


Let $X$ be the condition: for any  set $\{\gamma\}\subset
\{\Gamma\}$ of propagators and vertices
${\bf X}(\{\gamma\})$ is true if and only if  $\gamma$
constitutes a one-particle irreducible superficially
divergent subgraph of $\Gamma$.


\smallskip


\noindent Further, to
${\bf X}(\{\Gamma\}/\{\gamma\})$ we associate the  graph $\Gamma/\gamma$ 
which we obtain if we shrink $\gamma$ in $\Gamma$
to a point.\\



\noindent {\bf Prop.5}
$X$ is an orderly condition.\\
{\bf Proof:} $\#(P_X(\{\Gamma\}/\{\gamma\}))=\#P_{X_{\{\gamma\}}}(M)$.
Assume that  two elements of either of these two sets 
correspond to the same two subgraphs of $\Gamma$.
Then, if they are overlapping, nested or disjoint in
one of these two sets, they are so in the other as well. ~ $\Box$


\smallskip


We are interested in the set ${\cal P}_X(\{\Gamma\})$.
For a 1PI Feynman graph $\Gamma$,  $T_X(\{\Gamma\})$ is the 
forest  assigned to it in the sense of the previous
section.
In general, $T_X(\{\Gamma\})$ will be a sum
of rooted trees $T_X(p), p\in {\cal P}_X^{cit}(\{\Gamma\})$.
Note further that $\{\Gamma\}$ is irreducible
with respect to $X$ for all 1PI graphs $\Gamma$.



Define the depth $d(\Gamma)$ as 
\[
d(\Gamma):=n_v(T(\{\Gamma\})),
\]
as before.
This depth is well-defined for any Feynman graph.




Feynman diagrams without subdivergences thus have depth one,
as they correspond to the rooted tree $t_1$ decorated by the set $\{\Gamma\}$.


Each Feynman diagram has a well-defined depth and thus we have
a decomposition on the set of all Feynman graphs ${\cal FG}$,
\[
 {\cal FG}={\cal FG}^{[0]}\cup{\cal FG}^{[1]}\cup{\cal FG}^{[2]}\cup{\cal FG}^{[3]}
\cup\ldots 
\]
Here ${\cal FG}^{[0]}$ corresponds to superficially convergent graphs.
We are interested in graphs in ${\cal FG}^{[n]}$, $n\geq 1$.


To Feynman graphs of depth one we assign the rooted tree $t_1$,
decorated by the corresponding element of ${\cal FG}^{[1]}$.
The elements of this set furnish the set of primitive
elements of the Hopf algebra ${\cal H}_R({\cal FG}^{[1]})$ 
of decorated rooted trees.


The results of the previous section show that for each Feynman graph
$\Gamma\in {\cal FG}^{[k]}$, we find  a sum of
associated rooted tree
$T_\Gamma$ and a coproduct given by
\begin{equation}
\Delta(T_\Gamma)=1\otimes T_\Gamma
+T_\Gamma\otimes 1+\sum_{\gamma\subsetX \Gamma}T_\gamma\otimes
T_{\Gamma/\gamma}.\label{copfg}
\end{equation}
Here, $T_\Gamma$ is a sum of rooted trees with decorations
in ${\cal FG}^{[1]}$ and in $\cup_{i=2}^k U^{[i]}$, primitive elements
in the Hopf algebra of rooted trees, obtained
from Feynman graphs without subdivergences
(which, as said earlier in this paper, includes graphs which have other subgraphs
reduced to a point in them)
and iteratively constructed primitive elements in $U^{[i]}$
as described in the previous section. 


We will soon see 
explicit examples which indeed show that the so constructed
elements are indeed primitive, hence correspond
to analytic expressions without subdivergences.



At this stage, we can justify the notation of \cite{hopf} or
\cite{CK}, where vertices of rooted trees where decorated
by elements of ${\cal FG}^{[1]}$ \cite{CK}, 
which in the same spirit were used 
as letters
of parenthesized words in \cite{hopf}. In Prop.~2 we 
labelled each vertex $v$ of
$T(\{\Gamma\})$ by a subset $\{\gamma\}$ corresponding to a subgraph
$\gamma$ in our context.
$\gamma$ itself can have further subdivergences.
But then, condition $X$ and Prop.~2 ensure
that we could as well label vertices by elements of
$\gamma(v)/\gamma_v$, which correspond to graphs without
subdivergences.


Before we come to examples, let us first make sure that we really get 
Zimmermann's forest formula from (\ref{copfg}).
\subsection{Derivation of the forest formula}
To the coproduct (\ref{copfg})
belongs an antipode given by
\begin{equation}
S(T_\Gamma)=-T_\Gamma-\sum_{\gamma\subset\Gamma}S[T_\gamma]
T_{\Gamma/\gamma},
\label{anti}
\end{equation}
as one immediately checks.
As it is an antipode in a Hopf algebra of rooted trees, it can be written as a sum
over all cuts.
Set $T_\Gamma=\sum_i T_i$ 
for some decorated rooted trees $T_i$.  Then,
\begin{equation}
S(T_\Gamma)
=
\sum_i \sum_{\mbox{\tiny all cuts $C_i$
of $T_i$}}
(-1)^{n_{C_i}}
P^{C_i}(T_i)R^{C_i}(T_i).\label{anti2}
\end{equation}
Each such cut corresponds to a renormalization
forest, which we obtain if we box the corresponding
subgraphs in $\Gamma$, and vice versa \cite{CK}. 
\footnote{Note that we can easily identify
maximal forests here
(in the sense of renormalization theory), 
by using the $B_-$ operator on the trees $T_i$.}



Now, let $\phi$ be a ${\bf Q}$-linear  map which assigns to $T_\Gamma$
the corresponding Feynman integral. 
Further, let $\phi_R=\tau_R\circ\phi$ be a map
which assigns to $T_\Gamma$ the corresponding Feynman integral, evaluated under some
renormalization condition $R$. Hence, from $T_\Gamma$ we obtain
via $\phi$ a Feynman integral $\phi(T_\Gamma)$ in need of renormalization.
$\tau_R$ modifies  this Feynman integral, in a way such that
the result contains the divergent part of this integral.  
Essentially, $\tau_R$ extracts the divergences of 
$\phi(T_\Gamma)$ in a meaningful way \cite{Collins}.
Hence, as $\tau_R$ isolates divergences faithfully, 
differences $(id-\tau_R)(\phi(T_\Gamma))$
eliminate divergences in Feynman integrals.
Depending on the chosen renormalization scheme $R$, one can adjust
finite parts to fulfil renormalization conditions.
A detailed study of this freedom from the Hopf algebra
viewpoint can be found in \cite{new}.


We remind the reader of Sweedler's notation:
$\Delta(T_\Gamma)=\sum {T_\Gamma}_{(1)}\otimes {T_\Gamma}_{(2)}$.
Let us consider the antipode $\bar{e}(T_\Gamma)$ using Sweedler's notation:
\[
0=\bar{e}(T_\Gamma)=\sum S({T_\Gamma}_{(1)}){T_\Gamma}_{(2)}.
\]
This map vanishes identically. Note that
it can also be written as 
\[
m[(S\otimes id)\Delta(T_\Gamma)]\equiv\bar{e}(T_\Gamma)=0.
\]
But this map gives rise to a much more interesting map,
by composition with $\phi$,
\[
T_\Gamma\to \Gamma_R:=m[(S_R\otimes id)(\phi\otimes\phi)\Delta(T_\Gamma)].
\]
This map associates to the Feynman graph $\Gamma$ represented
by a unique sum of rooted trees the renormalized
Feynman integral $\Gamma_R$ \cite{hopf,CK}.


Its usual definition
\begin{equation}
\Gamma_R=(id-\tau_R)\left[\Gamma+\sum_{\gamma\subset\Gamma}Z_\gamma 
\Gamma/\gamma\right],\label{for}
\end{equation}
is recovered if we define
\begin{equation}
S_R[\phi(T_\gamma)]
\equiv Z_\gamma=
-\tau_R(\gamma)-\tau_R\left[\sum_{\gamma^\prime\subset\gamma}
Z_{\gamma^\prime}\gamma/\gamma^\prime\right].\label{usual}
\end{equation}
This map is derived from  the antipode
\begin{equation}
S[T_\gamma]=-T_\gamma-\sum_{\gamma^\prime\subset\gamma}S[T_{\gamma^\prime}]
T_{\gamma/\gamma^\prime}.
\end{equation}
Using $\phi$ to lift this to Feynman graphs, and using the freedom to alter
corresponding analytic expressions according to renormalization schemes $R$
one obtains (\ref{usual}).


Note that if one defines
\[
\phi_R=S_R\circ\phi\circ S,
\]
one has $S_R\circ \phi=\phi_R\circ S$ and hence
\begin{equation}
S_R[\phi(T_\gamma)]=\tau_R\left[
-\phi(T_\gamma)-\sum_{\gamma^\prime\subset\gamma}\phi_R(S[T_{\gamma^\prime}])
\phi(T_{\gamma/\gamma^\prime}).
\right]
\end{equation}
Hence, in accordance with \cite{hopf,CK} we find the $Z$-factor of a
graph $\gamma$ as derived from the antipode in the Hopf algebra
of rooted trees. Above, in (\ref{for}),
we recovered the original forest formula in its recursive
form. The non-recursive form is recovered with the same ease,
using (\ref{anti2}) instead of (\ref{anti}) \cite{hopf,CK}.
It reads
$$
\Gamma_R=(id-\tau_R)\left[\sum_i\sum_{\mbox{\tiny all normal cuts}}
(-1)^{n_{C_i}}\phi_{\tau_R}(P^{C_i}(T_i))\phi(R^{C_i}(T_i))\right]
$$
in a form which makes its finiteness obvious when we take into account
that the operation $\tau_R$ is defined to leave divergences
unaltered. $\phi_{\tau_R}(P^{C_i}(T_i))$ implies an iterative
application of $\tau_R$ as governed by the unique boxes
(the forests of classical renormalization theory) associated
with normal cuts \cite{CK}. 
Explicit realizations will be given elsewhere \cite{new},
as well as  a more detailed discussion of renormalization schemes,
renormalization group equations, operator product expansions and 
relations to cohomological properties of renormalizations.
\subsection{Examples}
We start with a simple example.
Let $v_1,v_2,\omega_1,\omega_2,\omega_3$ 
be the Feynman graphs indicated in Fig.(\ref{ol1}).
We then have, switching to a notation in PW's \cite{hopf},
\footnote{For example, in this notation $((v_1)\omega_1)$
corresponds to the tree $t_2$, with its root
decorated by $\omega_1$ and the other vertex decorated by $v_1$.
Decorated rooted trees and PW's on an alphabet of decorations
are in one-to-one correspondence \cite{CK}.}
\begin{eqnarray}
T_X(\omega_3) & = & ((v_1)\omega_1)+((v_2)\omega_2),\\
T_{\omega_3} & = & ((v_1)\omega_1)+((v_2)\omega_2)+(U_{\omega_3}),\\
(U_{\omega_3}) & = & (T_{\omega_3})-[((v_1)\omega_1)+((v_2)\omega_2)],\\
\Delta[T_{\omega_3}] & = & T_{\omega_3}\otimes e +e \otimes T_{\omega_3}
+(v_1)\otimes (\omega_1)+(v_2)\otimes\omega_2.
\end{eqnarray}
The graphs belong to ${\cal FG}^{[2]}$.
\bookfig{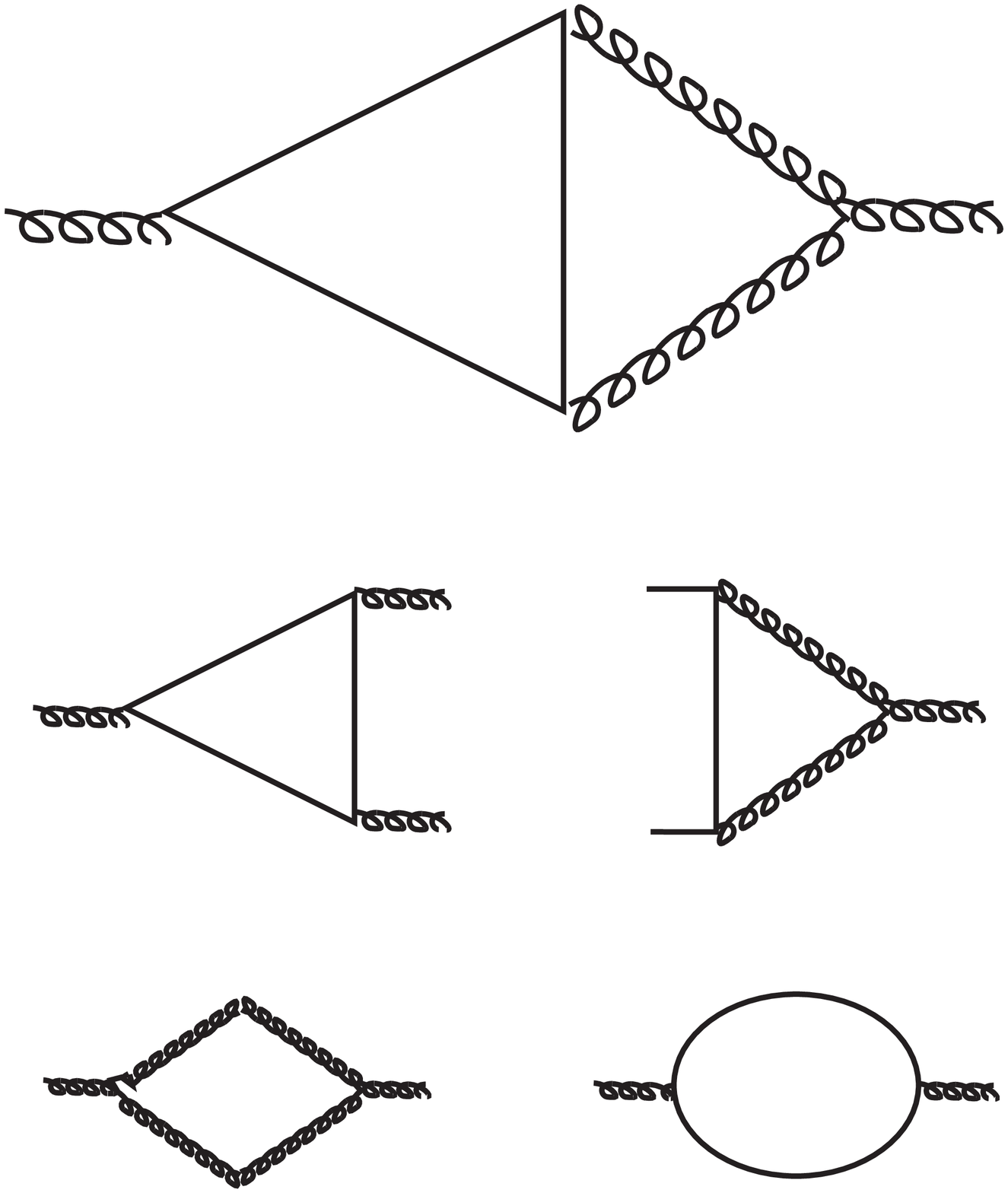}{${\cal FG}^{[2]}$}{ol1}{A graph from ${\cal FG}^{[2]}$
and its subgraphs. We read it as a graph in Yang-Mills theory in
four dimensions say, with straight lines being fermions.
In the first row, we see the graph $\omega_3$.
Below, we see its two subgraphs $v_1,v_2$
and in the bottom row we see the graphs
$\omega_1=\omega_3/v_1$ and $\omega_2=\omega_3/v_2$.}{6}


Note that $U_{\omega_3}$ gives us the skeleton corresponding
to this graph. It is a primitive element, and thus free of
subdivergences. And indeed, for any choice of momentum transfer and masses
in $v_i$, $\phi(U_{\omega_3})$ 
is an analytic expression free of subdivergences.
Any representation of  $T_\Gamma$ in terms
of Feynman integrals shows that the expressions corresponding
to such $U_\Gamma$ are free of subdivergences.
An instructive example is given in the appendix of \cite{CK},
where it is shown how graphs in $\phi^3$ theory explicitly
realize the results derived here on general grounds. Similar
results can be found in \cite{hopf,habil,BDK}.


Next, in Fig.(\ref{ol2}),
we consider  examples taken from ${\cal FG}^{[3]}$.
\bookfig{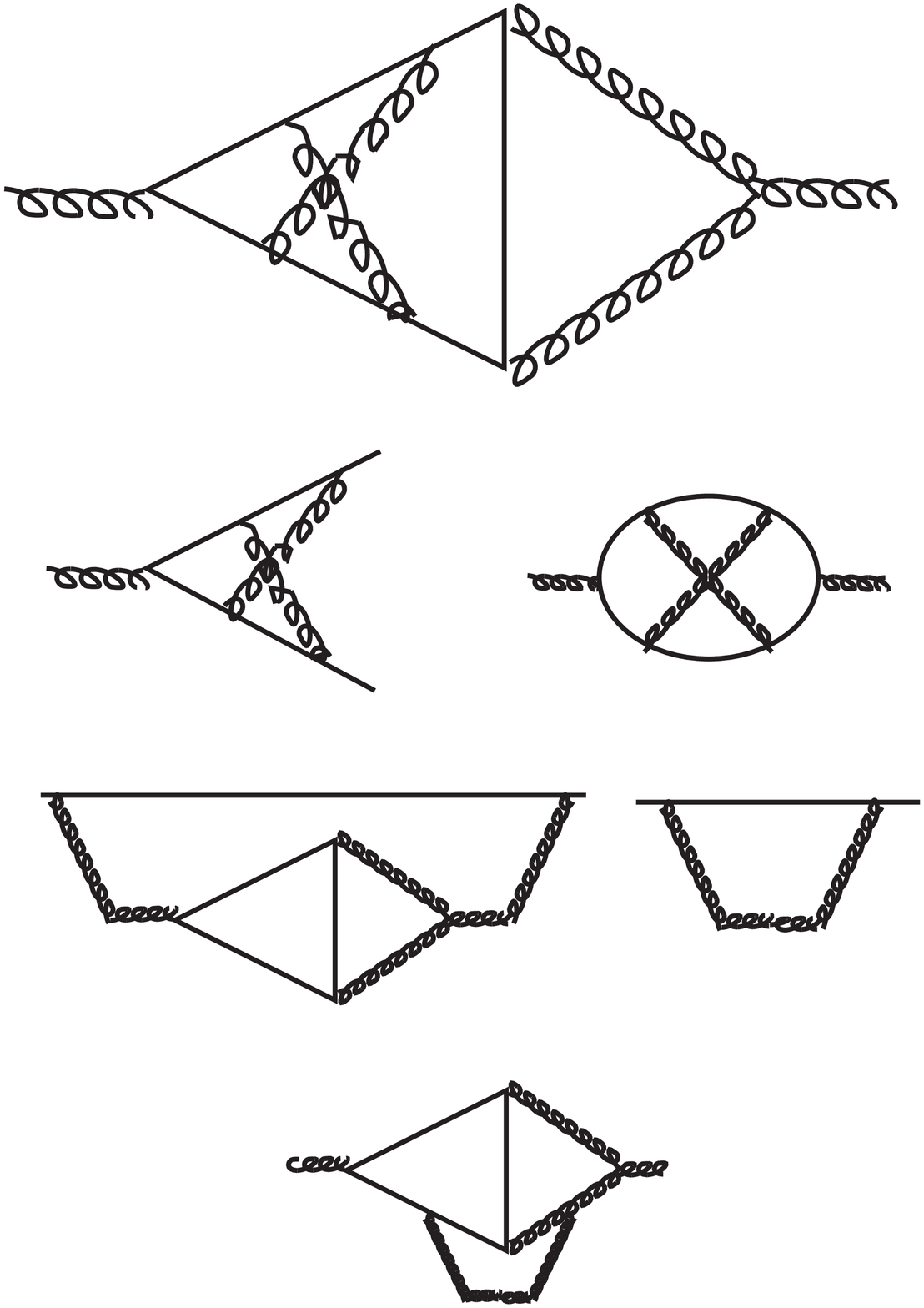}{${\cal FG}^{[3]}$}{ol2}{Graphs from ${\cal FG}^{[3]}$
and their subgraphs. At the top, we see the graph
$\omega_4$. Apart from the subgraphs in the previous figure,
we find two more subgraphs, the vertex $v_3$
and the self-energy $\omega_5=\omega_4/v_2$, both
given in the second row. In the third row,
we define the three-loop fermion self-energy $\Sigma_3$.
It involves the same subgraphs as before, plus a new
graph $\Sigma_1=\Sigma_3/\omega_3$. Finally,
at the bottom, we see the graph $\omega_6$.}{7}


This time, we find the following results
\begin{eqnarray}
T_X(\omega_4) & = & 
(((v_2)v_3)\omega_2)+(((v_3)v_1)\omega_1)+((v_3)(v_2)\omega_2),\\
\breve{T}(\omega_4) & = & ((v_2)U_{\omega_5})+((v_3)U_{\omega_3}),\\
T_{\omega_4} & = & 
(((v_2)v_3)\omega_2)+(((v_3)v_1)\omega_1)+((v_3)(v_2)\omega_2)\nonumber\\
 & & + ((v_2)U_{\omega_5})+((v_3)U_{\omega_3})+U_{\omega_4}.\\
\Delta(T_{\omega_4}) & = &
T_{\omega_4}\otimes e+e\otimes T_{\omega_4}\nonumber\\
 & & +2(v_2)\otimes ((v_3)\omega_2)+(v_3)\otimes 
((v_1)\omega_1)+(v_3)\otimes ((v_2)\omega_2)\nonumber\\
 & & +((v_2)v_3)
\otimes (\omega_2)+((v_3)v_1)\otimes
(\omega_1)+(v_3)(v_2)\otimes(\omega_2)
\nonumber\\
 & & +(v_2)\otimes(U_{\omega_5})+(v_3)\otimes(U_{\omega_3}).
\end{eqnarray}
Now,
\begin{eqnarray*}
 & & 2(v_2)\otimes ((v_3)\omega_2)+(v_3)\otimes ((v_1)\omega_1)
+(v_3)\otimes ((v_2)\omega_2)\\
 & & +(v_2)\otimes(U_{\omega_5})+(v_3)\otimes(U_{\omega_3})\\
 & & = (v_2)\otimes T_{\omega_5}+(v_3)\otimes T_{\omega_3},
\end{eqnarray*}
as
\begin{eqnarray*}
T_{\omega_5} & = & U_{\omega_5}-2((v_3)\omega_2),\\
T_{\omega_3} & = & U_{\omega_3}-((v_2)\omega_2)-((v_1)\omega_1).
\end{eqnarray*}
For the other graphs in Fig.(\ref{ol2})
we find
\[
T_{\Sigma_3}=(((v_1)\omega_1)\Sigma_1)+(((v_2)\omega_2)\Sigma_1)
+((U_{\omega_3})\Sigma_1),
\]
and
\[
T_{\omega_6}=(((v_2)v_1)\omega_1)+(((v_2)v_2)\omega_2)+
((v_2)U_{\omega_3}).
\]
We invite the reader to confirm that
the coproduct on these expressions has the desired form 
(\ref{natcop}).



Finally, Figs.(\ref{ol3},\ref{ol3b},\ref{ol3c})
shows how the transition from $T_X(\Gamma)$ to
$T_\Gamma$ is achieved in terms of surgery along edges.
We start with an example taken from $\phi^3$ theory in six
dimensions. We consider a quadratically divergent two-point
function as given in the figures.
Fig.(\ref{ol3}) gives $T_X(\Gamma)$. It consists
of six decorated rooted trees. In the figure, we give
the decorations not by primitive elements, but by full
subgraphs. The decoration by primitive elements is obtained, in accordance
with Prop.2, if we divide by the decorations at outgoing vertices.
That there are six decorated trees if a consequence
of the internal product structure of the graph: there is a subgraph
$\gamma_2$ with $\#({\cal P}_X^{cit}(\gamma_2))=2$,
and the complement graph $\Gamma/\gamma_2$ has
$\#({\cal P}_x^{cit}(\Gamma/\gamma_2))=3$.


Fig.(\ref{ol3b}) adds the terms for the transition
$T_X(\gamma)\to T_{\gamma}$. This is only  non-trivial for the
case that $\gamma$ is the indicated overlapping two-loop
two-point function $\gamma_2$. Finally, Fig.(\ref{ol3c})
shows the additional terms generated from the complement
graphs $\Gamma/\gamma$. 


\bookfig{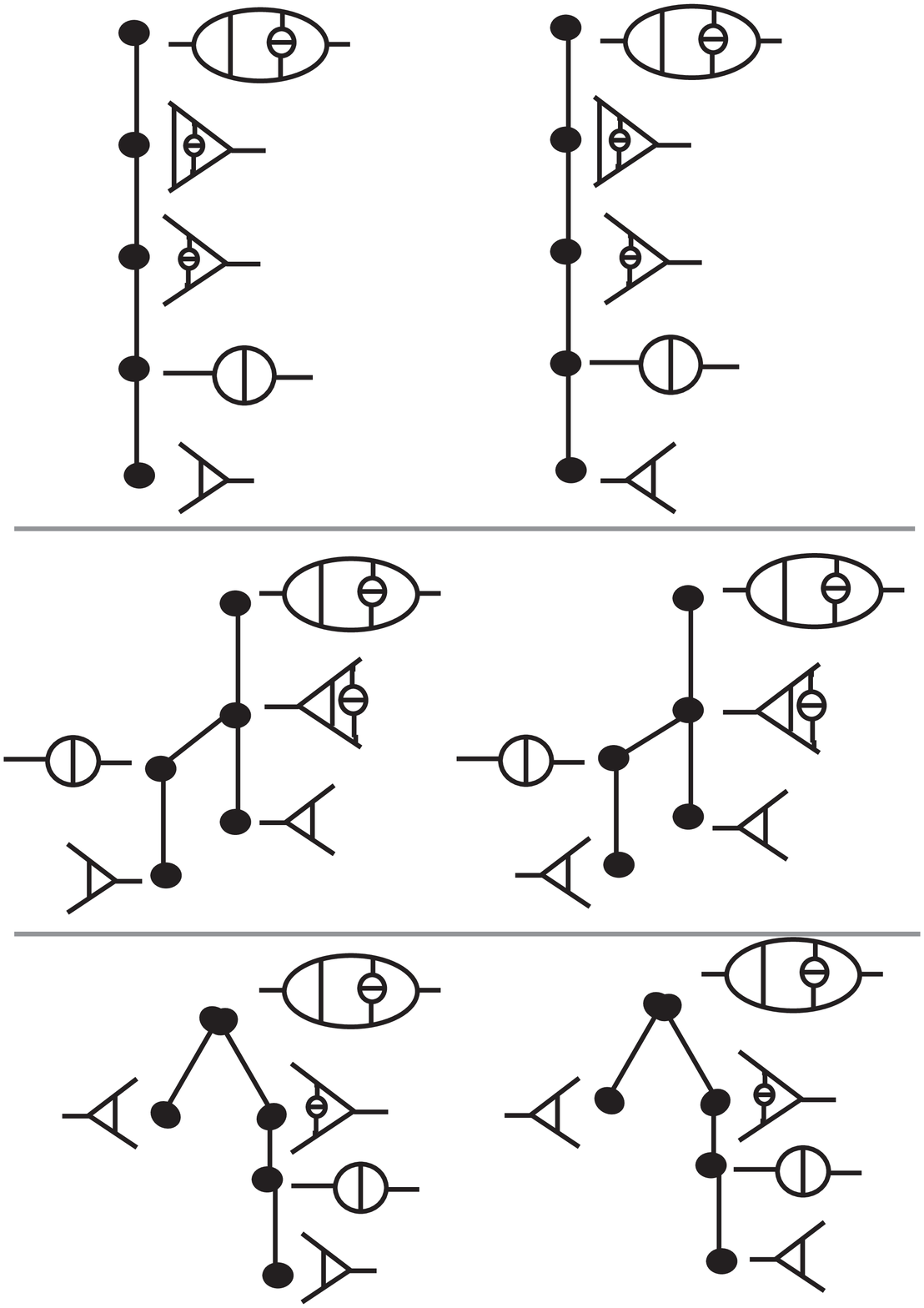}{Surgery at $T_X(\Gamma)$.}{ol3}{Surgery along edges delivers
the transition from $T_X(\Gamma)$ to $T_\Gamma$.
We first give $T_X(\Gamma)$, where $\Gamma$ is the five-loop
graph indicated at the roots. All six rooted trees
in this figure have to be added to give
$T_X(\Gamma)$.}{10}
\bookfig{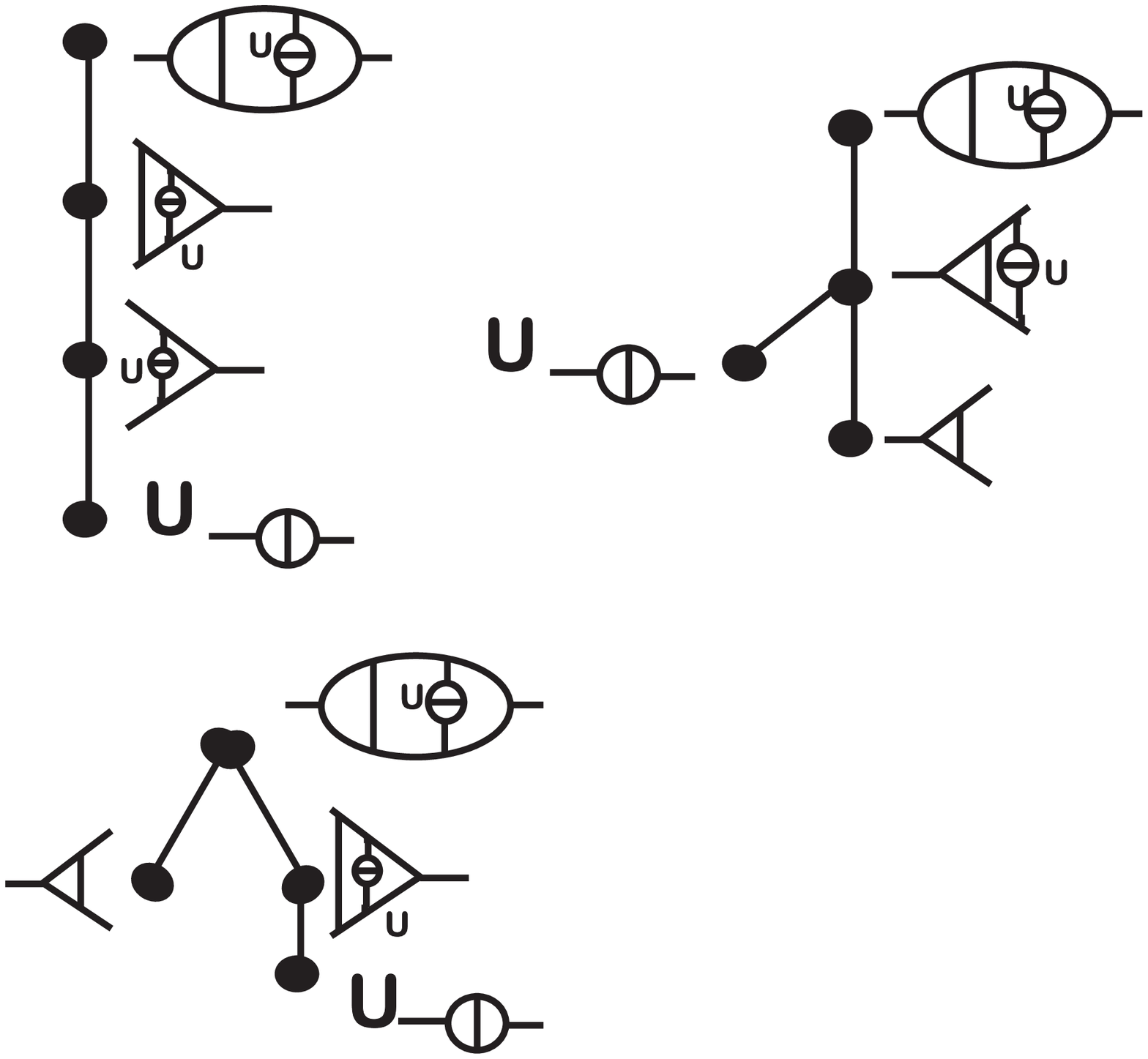}{Surgery at $T_X(\Gamma)$,II.}{ol3b}{
Now we add the results of replacing $T(\gamma)$ by $T_\gamma$.
}{7}
\bookfig{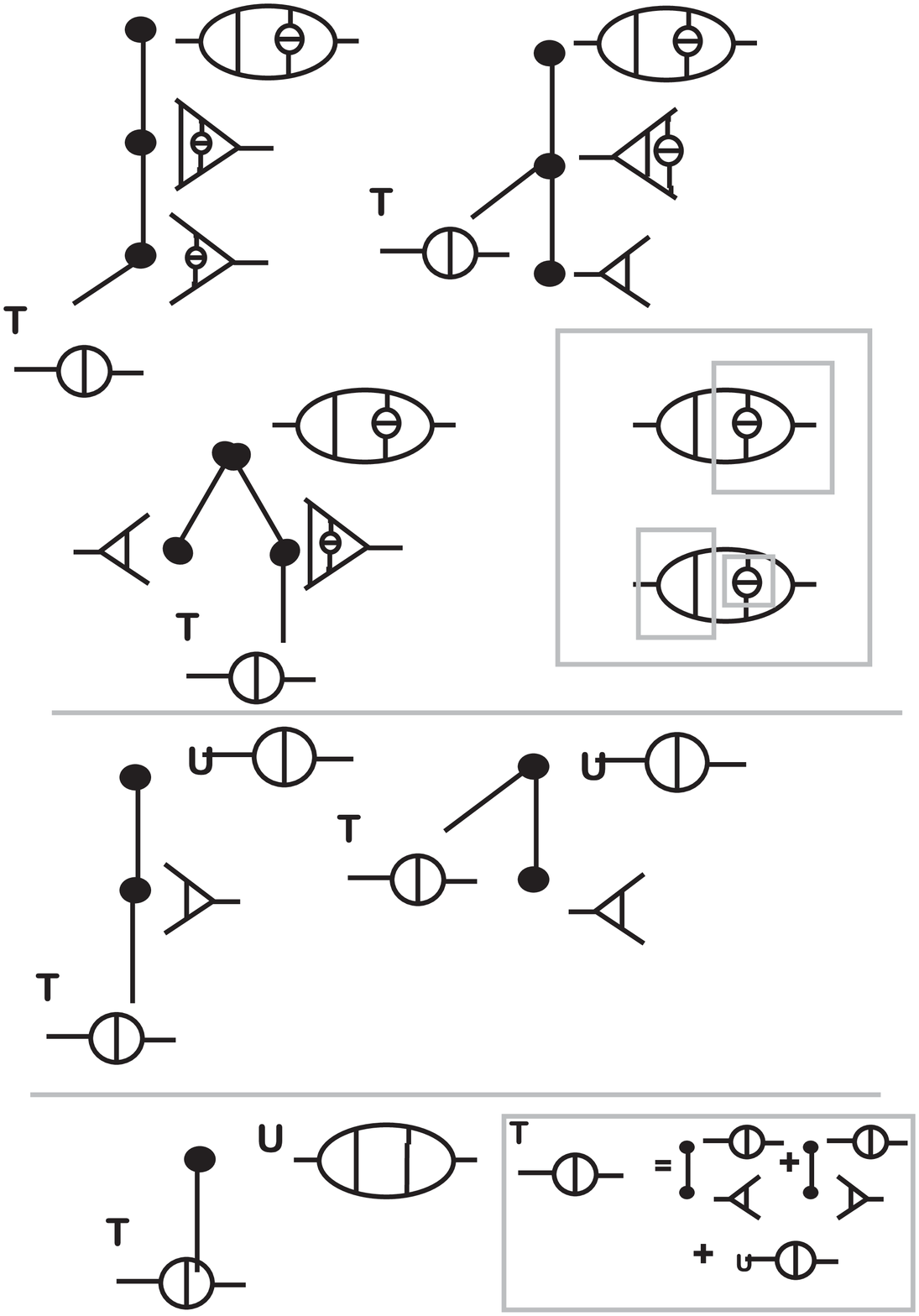}{Surgery at $T_X(\Gamma)$,III.}{ol3c}{
Finally, we construct the terms which achieve the transition
$T_X(\Gamma/\gamma)\to T_{\Gamma/\gamma}$. The first two rows, if we append 
the forest $T_\gamma$, give the terms of the previous two figures.
The second takes into account  the fact that in $\Gamma/\gamma$,
$\forall \gamma\in {\cal P}_X(\{\Gamma\})$,
we can find the element $\gamma_2=\Gamma/\gamma$ itself, by shrinking
three loops to this element of ${\cal FG}^{[2]}$. The inlay
in the first row indicates the graphs $\gamma$ which have to shrink.
Note that $\gamma$ is allowed to consist of disjoint 
graphs.
The last row takes
into account the primitive element $U_{\Gamma/\gamma_2}$.
The inlay defines  $T_{\gamma_2}$.
}{10}


Let us end this section with a few remarks concerning the various sorts
of overlapping divergences. Most prominent and most severe are overlapping
quadratic divergences, as one encounters typically
in (gauge)-boson propagators, let it be gauge theory
in four dimensions of $\phi^3$ theory in six, considered in the above examples.
Typically, the overlapping subdivergences are provided by vertex corrections,
and hence we have two sets which overlap.
Characteristically, the two overlapping subdivergences
can be eleminated by two derivatives with respect to an external
momentum, one for each of them.
This then generates new decorations of logarithmic degree
of divergence. An illuminating example for this situation is given
in the appendix of \cite{CK}. 


Overlapping degrees of divergences can come in other degrees
of divergence, and in other configurations. For example, in non-abelian
Yang-Mills theory one can have overlapping divergent Feynman graphs
with a logarithmic degree of divergence, where one has three sets which 
mutually overlap with each other.\footnote{A tetrahedron formed out of gluons,
with three external gluons coupling to three sides of the tetrahedron
which form a triangle is an appealing three-loop example only involving 
three-gauge-boson vertices.}


\section{Conclusions}
Starting from set-theoretic notions, we showed
how the forest formula underlying renormalization
theory is {\em ad initio}  derived from the Hopf algebra
of rooted trees. At the same time, we constructed a systematic way how to
obtain the skeleton expansion in any QFT,
given by elements $U_M$. We derived the original non-recursive
forest formula
of Zimmermann from the Hopf algebra of rooted trees, as well as the
recursive formulation.
The results of \cite{KW} are in full accordance with our results
and are a specification of the general result presented here.
 
 
Details for the practitioner
of calculational QFT are given elsewhere \cite{new},
including remarkable number-theoretic results
when investigating the role of the Connes-Moscovici
Hopf subalgebra in Feynman diagrams.


Some further remarks
are in order. 
\begin{itemize}
\item The methods developped in the first section are sufficiently general to be applied
to problems of operator product expansions
and asymptotic expansions, with applications
to OPE's already being established \cite{new}. 
Our approach being based
on set-theoretic considerations, the remaining challenge for
general asymptotic expansions is to find and
interpret sensible conditions $X$, and to identify the resulting
primitive elements.
\item The Hopf algebra of rooted trees has relations to
shuffle Hopf algebras \cite{MEH}. 
Shuffle products play a role when we start to study the action of the
symmetric group on
decorations. They appear naturally in the consideration
of the sub Hopf algebra generated by trees $B_+^k(e)$, which is
the Hopf algebra underlying Chen's iterated integral.
The Hopf algebra of rooted trees
has this algebra as a sub Hopf algebra. There are interesting generalizations
when we study shuffle algebras and iterated integrals 
from the viewpoint of the 
Hopf algebra of rooted trees \cite{new}.
Especially, the absence of a shuffle product for bare Green functions
in the presence of a remaining
convolution law points to interesting structures
lying ahead \cite{new}.
\item
The general set-theoretic set-up adopted in this paper allows
to study bare Green functions in $x$-space, and hence will allow
to study them as functions on configuration space
(which relies on tree-ordered boundaries in a natural manner,
see e.g.\cite{thurston} and references there).
This will hopefully reconcile early work on such functions
\cite{EG} with more recent developments.
\end{itemize}
\section*{Acknowledgements}
Let me first thank Raymond Stora for interest and discussions, and for
the ultimate motivation to write this paper. Also,
I thank him for carefully
proofreading an earlier version of this paper.
I very much enjoyed the opportunity
to discuss the intricate structures of the calculus of QFT
and to collaborate on its surprising relation to Noncommutative Geometry
with  Alain Connes. I also thank Alain for generous hospitality
on various occasions.
As usual, many thanks are due
to David Broadhurst for companionship in our longlasting exploration
of patterns and structures in QFT.
Support by a Heisenberg fellowship is gratefully acknowledged.


\begin{thebibliography}{99}
\bibitem{Zimm}
W.~Zimmermann, Comm.Math.Phys.{\bf 15} (1969) 208.
 \bibitem{hopf} 
D.~Kreimer, Adv.Theor.Math.Phys.{\bf 2}, 303 (1998);
q-alg/9707029.
\bibitem{habil}
D.~Kreimer, J.Knot Th.Ram.{\bf 6} 479 (1997); q-alg/9607022.
\bibitem{CK}
A.~Connes, D.~Kreimer, Comm.Math.Phys.{\bf 199}
203 (1998); hep-th/9808042. 
\bibitem{KW}
T.~Krajewski, R.~Wulkenhaar, {\em On Kreimer's Hopf algebra structure
of Feynman Graphs}, CPT-98/P.3639; hep-th/9805098, to appear
in Eur.Phys.J.{\bf C}.
\bibitem{Collins}
J.C.~Collins, {\em Renormalization}, Cambridge Univ.~Press (1984).
\bibitem{BDK}
D.J.~Broadhurst, R.~Delbourgo, D.~Kreimer,
Phys.Lett.{\bf B366} 421 (1996); hep-ph/9509296.
\bibitem{MEH}
M.E.~Hoffman, {\em Quasi-shuffle products}, preprint, 
to appear in J. Algebraic Combinatorics;\\
J.M.~Borwein, D.M.~Bradley, D.J.~Broadhurst, P.~Lisonek, 
{\em Combinatorial aspects of multiple zeta values,}
Electr.J.Comb.{\bf 5} (1998), R38. 
\bibitem{new}
D.J.~Broadhurst, D.~Kreimer, {\em Renormalization
automated by Hopf algebras}, hep-th/9810087;\\
D.~Kreimer, {\em Chen's iterated integral represents
the Operator Product Expansion}, hep-th/9901099;\\
R.~Delbourgo, D.~Kreimer, {\em Using Hopf algebras
to calculate Feynman diagrams}, in preparation.
\bibitem{thurston}
D.P.~Thurston, {\em
Integral Expressions for the Vassiliev Knot Invariants},
math/9901110.
\bibitem{EG}
H.~Epstein, V.~Glaser, R.~Stora,
{\em General Properties of the $n$-point
Functions in Local Quantum Field Theory},
in {\em Les Houches 1975}, Proceedings, {\em
Summer School On Structural Analysis Of Collision Amplitudes}, 
Amsterdam 1976, 5-93;\\
H. Epstein, V. Glaser, Ann.~Inst.~H.~Poincar\'e{\bf 19} 211 (1973).
\end{thebibliography}
\end{document}